\documentclass[a4paper,journal]{IEEEtran}
\usepackage[colorlinks,
            linkcolor=blue,
            anchorcolor=blue,
            citecolor=blue]{hyperref}
\usepackage{amsmath,amsfonts}
\usepackage{amsthm,amssymb,lipsum}
\usepackage{xcolor}

\usepackage{mathrsfs}
\usepackage[linesnumbered,ruled]{algorithm2e}
\SetKwInput{KwInput}{Input}                
\SetKwInput{KwOutput}{Output}              
\usepackage{array}
\usepackage[caption=false,font=normalsize,labelfont=rm,textfont=rm]{subfig}
\usepackage{algpseudocode}
\usepackage{textcomp}
\usepackage{stfloats}
\usepackage{booktabs}
\usepackage{url}
\usepackage{verbatim}
\usepackage{graphicx}
\usepackage{float}
\usepackage{cite}
\usepackage{enumitem}
\usepackage[numbers, square]{natbib}
\usepackage{flushend}
\usepackage{geometry}
\begin{document}

\title{Rate Splitting Multiple Access-Enabled Adaptive Panoramic Video Semantic Transmission}
\author{Haixiao Gao, \IEEEmembership{Student Member, IEEE}, Mengying Sun, \IEEEmembership{Member, IEEE}, Xiaodong Xu, \IEEEmembership{Senior Member, IEEE}, Shujun Han, \IEEEmembership{Member, IEEE}, Bizhu Wang, \IEEEmembership{Member, IEEE}, Jingxuan Zhang, \IEEEmembership{Member, IEEE}, Ping Zhang, \IEEEmembership{Fellow, IEEE}
\thanks{This paper is supported by the National Key R\&D Program of China No. 2020YFB1806900, in part by the China Postdoctoral Science Foundation under Grant 2023M740341, and in part by the Beijing Natural Science Foundation No. L232051. (\emph{Corresponding author: Xiaodong Xu.})

Haixiao Gao, Mengying Sun, Shujun Han, Bizhu Wang, and Jingxuan Zhang are with the State Key Laboratory of Networking and Switching Technology, Beijing University of Posts and Telecommunications, Beijing 100876, China (e-mail: haixiao@bupt.edu.cn, smy\_bupt@bupt.edu.cn, hanshujun@bupt.edu.cn, wangbizhu\_7@bupt.edu.cn, zhangjingxuan@bupt.edu.cn).

Xiaodong Xu and Ping Zhang are with the State Key Laboratory of Networking and Switching Technology, Beijing University of Posts and Telecommunications, Beijing 100876, China, and also with the Department of Broadband Communication, Peng Cheng Laboratory, Shenzhen 518066, Guangdong, China (e-mail: xuxiaodong@bupt.edu.cn, pzhang@bupt.edu.cn).}
}



\maketitle
\begin{abstract}
In this paper, we propose an adaptive panoramic video semantic transmission (APVST) framework enabled by rate splitting multiple access (RSMA). The APVST framework consists of a semantic transmitter and receiver, utilizing a deep joint source-channel coding structure to adaptively extract and encode semantic features from panoramic frames. To achieve higher spectral efficiency and conserve bandwidth, APVST employs an entropy model and a dimension-adaptive module to control the transmission rate. Additionally, we take weighted-to-spherically-uniform peak signal-to-noise ratio (WS-PSNR) and weighted-to-spherically-uniform structural similarity (WS-SSIM) as distortion evaluation metrics for panoramic videos and design a weighted self-attention module for APVST. This module integrates weights and feature maps to enhance the quality of the immersive experience. Considering the overlap in the field of view when users watch panoramic videos, we further utilize RSMA to split the required panoramic video semantic streams into common and private messages for transmission. We propose an RSMA-enabled semantic stream transmission scheme and formulate a joint problem of latency and immersive experience quality by optimizing the allocation ratios of power, common rate, and channel bandwidth, aiming to maximize the quality of service (QoS) scores for users. To address the above problem, we propose a deep reinforcement learning algorithm based on proximal policy optimization (PPO) with high efficiency to handle dynamically changing environments. Simulation results demonstrate that our proposed APVST framework saves up to 20\% and 50\% of channel bandwidth compared to other semantic and traditional video transmission schemes, respectively. Moreover, our study confirms the efficiency of RSMA in panoramic video transmission, achieving performance gains of 13\% and 20\% compared to NOMA and OFDMA.
\end{abstract}


\section{Introduction}
\IEEEPARstart{I}{n} the vision of 6G, immersive communication will be a key application scenario where users are invited to watch 360-degree panoramic videos and interact with the virtual world \cite{itu}, \cite{immersive_communication}. To ensure the quality of service (QoS), it needs to provide low-latency and high-quality video transmission with less than $20\,$ms \cite{modeling_tradeoff}, \cite{scalable_video}. However, for a 4K video ($3840\times1920$), the user can actually view the image at a resolution of approximately $960\times540$ because of the limitations of the field of view (FoV) \cite{apvst}. Therefore, it needs to transmit high-resolution panoramic video to achieve a greater immersive experience, which leads to a dramatic increase in the amount of data transmitted. 

Current panoramic video transmission schemes are designed based on the traditional wireless communication system, and these schemes employ bitrate selection, resource allocation, and other strategies to achieve low-latency and high transmission performance metrics \cite{transcoding_enabled}. In the context of multiple-input multiple-output (MIMO) and orthogonal frequency-division multiple access (OFDMA), Guo \emph{et al}. constructed a MIMO-OFDMA system for adaptive panoramic video transmission \cite{OFDMA_VR}. This system is capable of adaptively selecting the quality of tiles, transmission power, and beamforming based on the current environment. Using non-orthogonal multiple access (NOMA), Li \emph{et al}. proposed a NOMA-assisted VR content transmission scheme aimed at minimizing the cost-oriented towards QoS \cite{NOMA_VR}. With the introduction of rate-splitting multiple access (RSMA), which is noted for its flexibility, it is increasingly being applied in various scenarios. Mao \emph{et al}., in their study on RSMA, confirmed that it achieves higher spectral efficiency and system throughput compared to orthogonal multiple access (OMA), space division multiple access (SDMA), and NOMA \cite{RSMA}.

The core concept of RSMA is to split user messages into common and private messages, enabling the capability to partially decode interference and to treat part of the interference as noise. This contrasts with the extreme interference management strategies used in SDMA and NOMA. RSMA can automatically switch between SDMA and NOMA by adjusting the power and content of common and private streams. When the interference level is low, RSMA can operate closer to the mode of SDMA, where it allocates most resources to private data streams, leveraging the advantages of spatial separation. It minimizes inter-user interference through directed beams and spatial multiplexing, thereby enhancing the system's spectral efficiency and throughput. Conversely, when the interference level is high, RSMA tends to adopt a strategy similar to NOMA, utilizing different power levels to distinguish signals between users \cite{RSMA}. For the transmission of panoramic videos, RSMA can flexibly adjust the allocation ratios of common and private messages within and outside the user's FoV, thereby reducing resource usage while ensuring the quality of user experience. Hieu \emph{et al}. developed a joint communication and computational optimization problem based on RSMA, grouping users according to their FoV and employing multicast for VR video streams \cite{vr_meet_rsma}. Huang \emph{et al}. proposed a VR streaming system assisted by intelligent reflective surfaces and RSMA \cite{ris_vr_rsma}, utilizing deep reinforcement learning (DRL) algorithms to optimize IRS phase shifts, RSMA parameters, and beamforming vectors, thus enhancing QoS.

However, while existing systems that employ RSMA enhance spectral efficiency, they also potentially increase the data transmission volume by splitting the original panoramic video streams. Against this backdrop, semantic communication has garnered widespread attention as a key enabling technology for 6G \cite{semantic_communication1}, \cite{semantic_communication2}, \cite{semantic_communication3}. It significantly reduces the data transmission requirement by extracting and compressing information at the semantic level, thus enhancing the efficiency of panoramic video transmission. Compared to traditional bit-level error-free transmission, semantic communication effectively overcomes the ``cliff effect'' caused by declines in signal-to-noise ratio (SNR) \cite{semantic_cliff_effect}. This approach not only ensures the accurate delivery of the core content and meaning of information but also enhances communication robustness in unstable or complex network environments. Unlike traditional bit-oriented communication, semantic communication aims to pursue higher ``semantic fidelity'' rather than error-free transmission of symbols. By applying neural networks, artificial intelligence enables the extraction and recovery of semantic features of information, thus reducing the amount of data transmitted and achieving semantic-level compression of image, text, speech, and other types of information \cite{semantic_communication4}, \cite{semantic_communication5}. The application of this technology is expected to resolve challenges associated with simultaneous online interactions and high-quality video transmission for multiple users, thus opening new frontiers in immersive communication \cite{semantic_communication_immersive_communication}.

In order to achieve efficient transmission of video, there has been a lot of research on how to employ neural networks for video transmission. Li \emph{et al}. build the deep contextual video compression (DCVC) network \cite{DCVC}, which extracts rich high-dimensional contextual information by the correlation between frames to transmit videos. After that, Wang \emph{et al}. of \cite{DVST} proposed the deep video semantic transmission (DVST) network, which takes the semantic communication into account. The DVST network achieves semantic transmission of video by nonlinear transforms and deep joint source-channel coding (Deep JSCC) under the guidance of the entropy model. In this paper, to transmit the panoramic video, we first apply equirectangular projection (ERP) to project spherical video on a flat surface. However, due to the characteristics of the ERP \cite{End_to_End_Optimized_360_Image_Compression}, there would be a large information redundancy if panoramic videos are transmitted directly by the DCVC and DVST networks. At the same time, it will cause a decrease in the panoramic video evaluation metrics weighted-to-spherically-uniform peak signal-to-noise ratio (WS-PSNR)\cite{WS_PSNR} and weighted-to-spherically-uniform structural similarity (WS-SSIM) \cite{WS_SSIM} which mean the quality of immersive experience. 

To improve the transmission efficiency of panoramic videos and bring better immersive experience quality, a more efficient RSMA-enabled semantic communication framework is needed. Overall, we will explore how to utilize RSMA to split semantic video stream messages, thereby further enhancing the efficiency of panoramic video transmission for users. The contributions of this paper can be summarized as follows:
\begin{itemize}
    \item We propose an RSMA-enabled adaptive panoramic video semantic transmission framework. This framework extracts semantic features by leveraging correlations between consecutive video frames. For such a framework, considering the user's FoV, RSMA is employed to efficiently split the semantic stream into common and private messages, thereby facilitating effective transmission of panoramic videos.
    \item We design a semantic transmitter and receiver for efficient adaptive panoramic video semantic transmission, i.e., APVST. The proposed APVST is based on the Deep JSCC structure and attention mechanism to adaptively extract semantic features from panoramic frames and achieve variable-length encoding of semantic information. To achieve higher WS-PSNR and WS-SSIM, we propose a weight attention (WA) module to learn the correlation between weights and semantic features for higher-quality panoramic video transmission. Since ERP causes the information redundancy on different latitudes, we propose a latitude adaptive module which can reduce the bandwidth for panoramic video transmission by combining it with the entropy model.
    \item We propose a mapping method of original FoV information to the semantic level and have validated the feasibility of this mapping through empirical testing. Furthermore, we propose an RSMA-enabled semantic stream transmission scheme, which is formulated as a joint optimization problem of latency and immersive experience quality. The optimization goal of this problem is to maximize users' QoS scores, which are defined as a weighted combination of latency score and immersive experience quality score. To address this problem, we formulate it as a Markov Decision Process (MDP) and employ the PPO-based algorithm to guide policy towards enhancing system efficiency and users' QoS. This approach allows us to dynamically adjust the transmission strategy to accommodate environmental changes and varying demands.
    \item We conducted extensive experimental validation of the proposed scheme. Simulation results indicate that our APVST framework can save 20\% and 50\% in channel bandwidth cost compared to other semantic and traditional video transmission schemes, respectively. Moreover, compared to NOMA and OFDMA, RSMA-enabled APVST scheme achieves gains of 13\% and 20\% in the transmission of semantic streams for panoramic videos.
\end{itemize}

\begin{figure}[t]
\centering
\includegraphics[width=85mm]{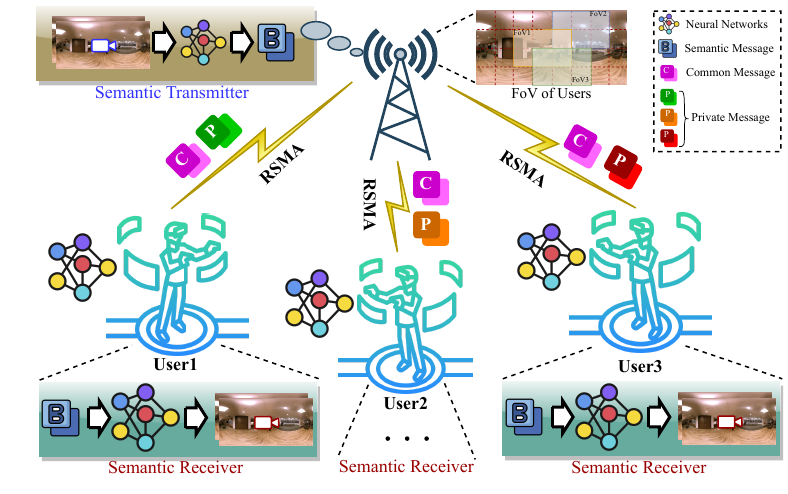}
\caption{System model of RSMA-enabled adaptive panoramic video semantic transmission.}
\label{system model}
\end{figure}

The rest of this paper is organized as follows. The system model and problem formulation are described in Section \ref{system model and problem formulation}. Section \ref{internal structure of apvst} introduces the internal structures of APVST as well as its training procedure. Details of our PPO-based approach for RSMA-enabled semantic stream transmission are described in Section \ref{DRL section}. Experimental Results are analyzed in Section \ref{section experiment}. Conclusions are finally drawn in Section \ref{conclusion}.

\begin{figure*}[htbp]
\centering
\includegraphics[width=178mm]{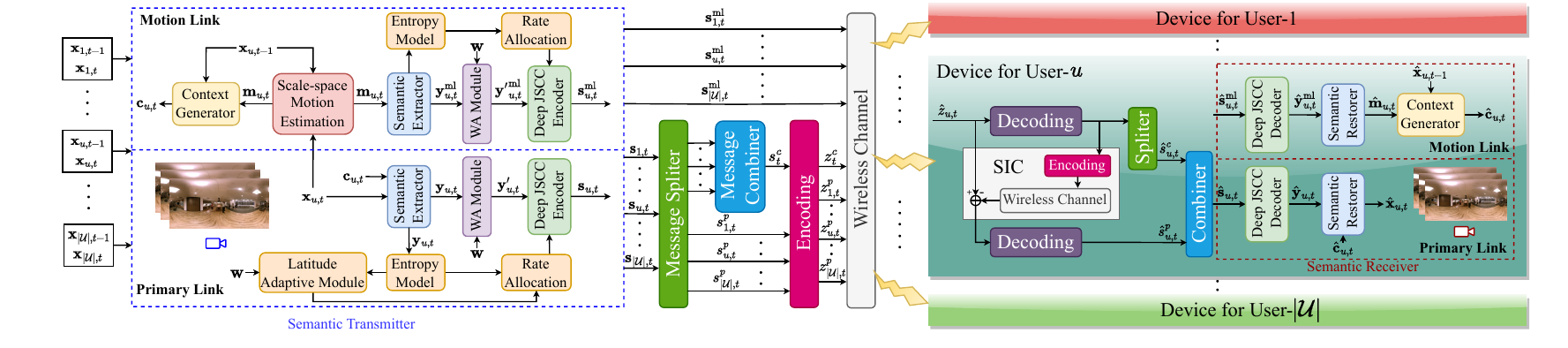}
\caption{The framework of RSMA-enabled APVST.}
\label{framework of system model}
\end{figure*}

\section{System Model and Problem Formulation} \label{system model and problem formulation}
We consider a scenario involving the downlink transmission of a panoramic video, which is facilitated by RSMA and based on semantic communication technology. In this scenario, a panoramic video is transmitted to \( |\mathcal{U}| \) users via a single base station (BS) with \( M \)-antennas. Our proposed system model is illustrated in Fig.~\ref{system model}. The set of users is denoted by \( \mathcal{U}=\{1,2,\ldots,u,\ldots,|\mathcal{U}|\} \). Given that the transmitted content is video and considering the variability of the environment, a time period is divided into multiple time slots, represented by \( \mathcal{T}=\{1,2,\ldots,t,\ldots\} \). The panoramic video frame requested by user $u$ at time slot $t$ is denoted by \( \mathbf{x}_{u,t} \), which is transformed into the semantic coding information \( \mathbf{s}_{u,t} \) at the BS through semantic extraction and Deep JSCC encoding. Subsequently, the semantic coding information is split into common and private streams by RSMA, enabling rate adaptation and facilitating the efficient downlink transmission of semantic information. At the receiver, the user's head-mounted display (HMD) demodulates the semantic information stream via successive interference cancellation (SIC) and achieves the recovery of the panoramic video frame \( \mathbf{\hat{x}}_{u,t} \) through Deep JSCC decoding and semantic restoration.

\subsection{End-to-end Transmission Process}
The proposed RSMA-enabled adaptive panoramic video semantic transmission framework is shown in Fig.~\ref{framework of system model}. We refer to the semantic transmitter and semantic receiver together as APVST, which consists of the motion link and the primary link. Referring to \cite{DCVC}, \cite{DVST}, based on the rich correlation information between current and reference panoramic frames, the APVST estimates the motion information and transmits it to the receiver. Based on the motion information and reference panoramic frame, the context generator extracts the context to enable more efficient panoramic video transmission. The primary link adaptively converts the source frames into potential representations in the semantic feature domain with the context. To enable wireless transmission, the semantic information is encoded at variable length by the Deep JSCC encoder, while the code rate is jointly guided by the learnable latitude adaptive module and entropy model. 

As shown in Fig.~\ref{framework of system model}, the motion link estimates the motion between $\mathbf{m}_{u,t}$ the current and reference panoramic frames $\mathbf{x}_{u,t}$ and ${\mathbf{x}}_{u,t-1}$, which can be expressed as
\begin{equation}
\mathbf{m}_{u,t}=f_\text{ME}\left(\mathbf{x}_{u,t},{\mathbf{x}}_{u,t-1}\right),\label{eq1}
\end{equation}
where $f_\text{ME}$ denotes the function of Scale-space Motion Estimation. In order for the receiver to get the context, semantic extraction is performed on the $\mathbf{m}_{u,t}$ to get the motion semantic feature $\mathbf{y}_{u,t}^\text{ml}$. Immediately after that, it is encoded as codeword sequence $\mathbf{s}_{u,t}^\text{ml}$ and fed into the channel for transmission. This process can be expressed as
\begin{equation}
\mathbf{y}_{u,t}^\text{ml}=f_\text{SE}^\text{ml}\left(\mathbf{m}_{u,t}\right)\ \text{and} \ \mathbf{s}_{u,t}^\text{ml}=f_\text{JE}^\text{ml}\left({\mathbf{y^\prime}}_{u,t}^\text{ml}\right),\label{eq2}
\end{equation}
where $f_\text{SE}^\text{ml}$ and $f_\text{JE}^\text{ml}$ denote the function of the semantic extractor and Deep JSCC encoder in the motion link, respectively. As the input to the encoder, the weighted spatial attention feature  ${\mathbf{y^\prime}}_{u,t}^\text{ml}$ is obtained by the WA module of the motion link. Corresponding to the transmitter, the received codeword sequence ${\hat{\mathbf{s}}}_{u,t}^\text{ml}$ is decoded into motion semantic feature ${\hat{\mathbf{y}}}_{u,t}^\text{ml}$ at the receiver. At the same time, the semantic restorer is used to recover motion information ${\hat{\mathbf{m}}}_{u,t}$, i.e,
\begin{equation}
{\hat{\mathbf{y}}}_{u,t}^\text{ml}=f_\text{JD}^\text{ml}\left({\hat{\mathbf{s}}}_{u,t}^\text{ml}\right)\ \text{and} \ {\hat{\mathbf{m}}}_{u,t}=f_\text{SR}^\text{ml}\left({\hat{\mathbf{y}}}_{u,t}^\text{ml}\right),\label{eq3}
\end{equation}
where $f_\text{JD}^\text{ml}$ and $f_\text{SR}^\text{ml}$ denote the function of the Deep JSCC decoder and semantic restorer in the motion link, respectively. In order to obtain context $\mathbf{c}_{u,t}$ and $\mathbf{\hat{c}}_{u,t}$ at the transmitter and receiver, respectively, the motion information ${\mathbf{m}}_{u,t}$ and ${\hat{\mathbf{m}}}_{u,t}$ is fed into the context generator by combining it with the reference frames ${\mathbf{x}}_{u,t-1}$ and ${\hat{\mathbf{x}}}_{u,t-1}$. This process can be expressed as
\begin{equation}
\mathbf{c}_{u,t}\!=\!f_\text{CG}\left({\mathbf{m}}_{u,t},{\mathbf{x}}_{u,t-1}\right) \text{and} \ \mathbf{\hat{c}}_{u,t}\!=\!f_\text{CG}\left({\hat{\mathbf{m}}}_{u,t},{\hat{\mathbf{x}}}_{u,t-1}\right),\label{eq4}
\end{equation}
where $f_\text{CG}$ denotes the function of the context generator.

After the motion link generates $\mathbf{c}_{u,t}$ and ${\hat{\mathbf{c}}}_{u,t}$, the primary link automatically learns the correlation between the current frame $\mathbf{x}_{u,t}$ and $\mathbf{c}_{u,t}$ to remove the information redundancy. The semantic feature map $\mathbf{y}_{u,t}$ is extracted on the basis of $\mathbf{c}_{u,t}$ and encoded into the corresponding codeword sequence $\mathbf{s}_{u,t}$ by Deep JSCC encoding, i.e,
\begin{equation}
\mathbf{y}_{u,t}=f_\text{SE}\left(\mathbf{x}_{u,t}\middle|\mathbf{c}_{u,t}\right)\ \text{and} \ \mathbf{s}_{u,t}=f_\text{JE}\left(\mathbf{y}_{u,t}^\prime\right),\label{eq5}
\end{equation}
where $f_\text{SE}$ and $f_\text{JE}$ denote the function of the semantic extractor and Deep JSCC encoder in the primary link, respectively. As the input to the encoder, the weighted spatial attention feature $\mathbf{y}_{u,t}^\prime$ is obtained by the WA module. 

In order to achieve efficient transmission of panoramic video semantic information, we will further improve users' QoS based on RSMA. As illustrated in Fig.~\ref{system model}, there are overlapping and non-overlapping regions of the content viewed by the users due to their different FoVs. The encoded semantic information streams can be decomposed into common and private streams by RSMA. The former corresponds to overlapping regions and the latter corresponds to non-overlapping regions. However, due to the complexity of the neural network, the video frames have scattered the feature information to various regions during the process of semantic extraction and Deep JSCC encoding. This also directly leads to the fact that the user's FoV region cannot be directly mapped to the corresponding semantic coding map, which reduces the interpretability of RSMA in semantic stream transmission. In the next, the feasibility of the proposed mapping scheme will be verified by experiments.



In particular, the process of masking the semantic encoding sequence $\mathbf{s}_{u,t}$ corresponding to the panoramic frame $\mathbf{x}_{u,t}$ using multiplication coefficients is as follows:
\begin{itemize}
    \item The coefficients within the FoV region are set to 1, indicating that no changes are made to the information within that area.
    \item The mask coefficients $\rho$ outside the FoV region are set to a negative number.
    \item The coefficient matrix is subjected to consecutive average pooling downsampling to obtain the mask matrix.
    \item The mask matrix is vectorized to obtain the mask vector.
    \item The mask vector is multiplied by the semantic encoding sequences $\mathbf{s}_{u,t}$ of the primary link and the semantic encoding sequences $\mathbf{s}^\text{ml}_{u,t}$ of the motion link, respectively.
    \item  The reconstructed visualization of the mask and the non-mask area of the panoramic frame is observed.
\end{itemize}

\begin{figure*}[!t]
    \centering
        \subfloat[$\rho\!=\!1, \text{PNSR}_{\text{FoV}}\!=\!36.9$]{\includegraphics[width=125pt]{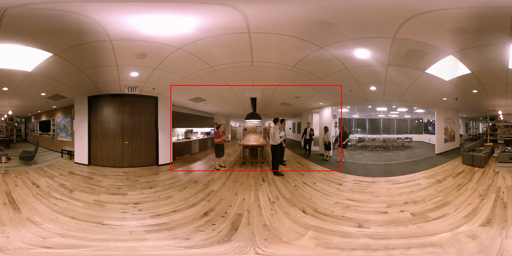}}%
    \hfill
        \subfloat[$\rho\!=\!-5, \text{PNSR}_{\text{FoV}}\!=\!32.6$]{\includegraphics[width=125pt]{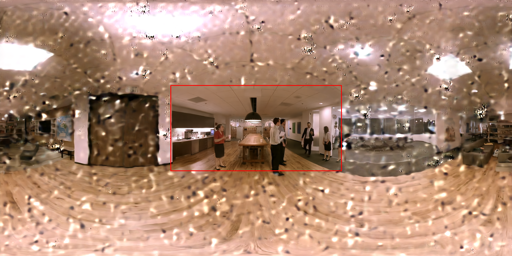}}
    \hfill
        \subfloat[$\rho\!=\!-6, \text{PNSR}_{\text{FoV}}\!=\!29.8$]{\includegraphics[width=125pt]{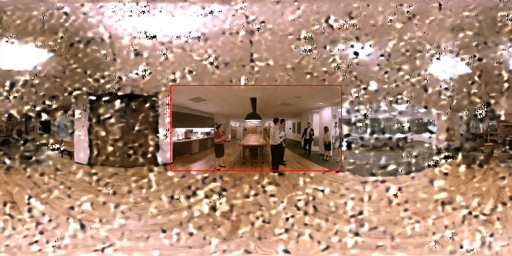}}%
    \hfill
        \subfloat[$\rho\!=\!-7, \text{PNSR}_{\text{FoV}}\!=\!26.5$]{\includegraphics[width=125pt]{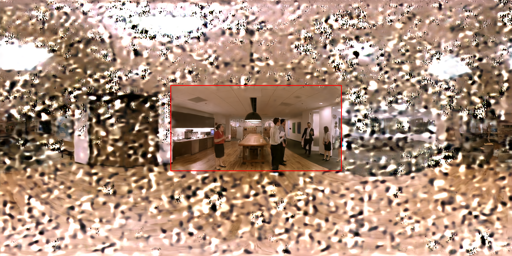}}%
    \vspace{-0.23cm}
    \hfill
        \subfloat[$\rho\!=\!-9, \text{PNSR}_{\text{FoV}}\!=\!21.1$]{\includegraphics[width=125pt]{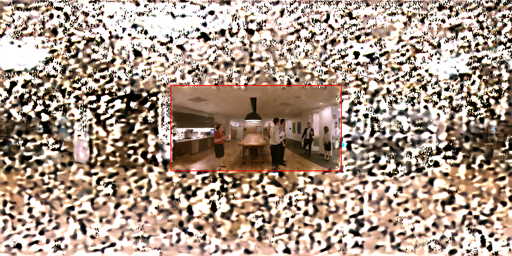}}%
    \hfill
        \subfloat[$\rho\!=\!-11, \text{PNSR}_{\text{FoV}}\!=\!18.2$]{\includegraphics[width=125pt]{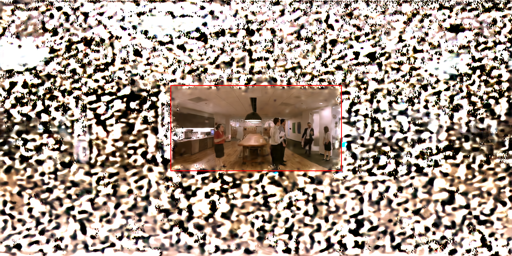}}%
    \hfill
        \subfloat[$\rho\!=\!-13, \text{PNSR}_{\text{FoV}}\!=\!16.5$]{\includegraphics[width=125pt]{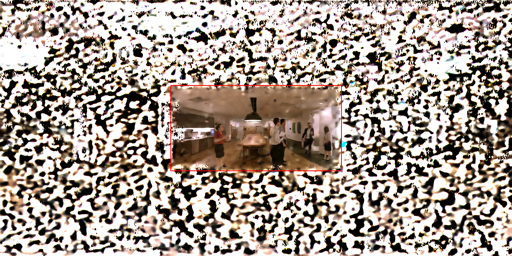}}%
    \hfill
        \subfloat[$\rho\!=\!-15, \text{PNSR}_{\text{FoV}}\!=\!15.1$]{\includegraphics[width=125pt]{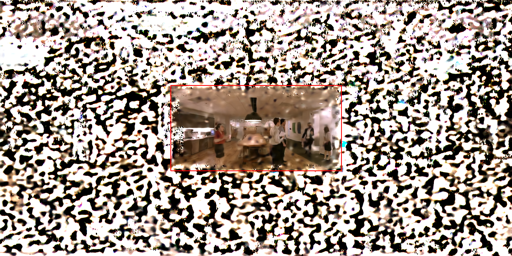}}%
    \caption{The reconstructed visualization of the mask and the non-mask area of the panoramic frame. $\text{PNSR}_{\text{FoV}}$ is the PSNR of the FoV area.}
    \label{fov experiment}
\end{figure*}

As illustrated in Fig.~\ref{fov experiment}, we assume the current FoV of the user is the area within the red box. It can be observed that the non-mask area (i.e., the FoV area) retains the original information of the panoramic frame, whereas the mask area loses most of the panoramic frame information. This indicates that most of the information within the FoV area is preserved at the corresponding position in the semantic encoding sequence after Deep JSCC encoding. Moreover, as $\rho$ is set increasingly smaller, the image restoration quality within the FoV area deteriorates. This also demonstrates that during the semantic communication process, a small portion of information within the FoV area exists in the semantic encoding sequence corresponding to the masked area.

Based on the above experiments and analysis, the mapping from the panoramic frame level to the semantic level of FoV information can be completed by setting the weight coefficient matrix and subjecting it to downsampling and vectorization. For example, by setting the weight values within the FoV area to 1 and the weight values in other areas to $\xi$, an FoV information map $\mathbf{I}_{u,t}$ can be obtained, where $\xi$ is an adjustable smaller weight value. By subjecting $\mathbf{I}_{u,t}$ to consecutive average pooling downsampling and vectorization, the semantic-level FoV information map $\mathbf{I}_{u,t}^s$ can be obtained.

Upon completing the mapping of FoV information from panoramic videos to semantic encoding sequences, the BS strategically determines which data should be incorporated into a common message and which should be categorized under private messages. Through the meticulous processing of panoramic frames' codeword sequences  $\mathbf{s}_{u,t}$ utilizing specially designed message separators and combiners, one common semantic message stream  $s_{t}^c$ and $|\mathcal{U}|$ unique private semantic message streams $s_{u,t}^p$ are formed.  The concepts of ``common" and ``private" semantic message streams reflect the overlapping and independent characteristics of the tile content viewed by users at the semantic level. Subsequently, the common semantic message stream for all users is encoded using a corresponding codebook, while the private semantic message streams for all users are encoded using the same codebook, respectively. The coding message streams $\mathbf{z}_t=\left[z_{t}^c,z_{1,t}^p,\ldots,z_{u,t}^p,\ldots,z_{|\mathcal{U}|,t}^p\right]^T$ are created from semantic message streams, where $z_{t}^c$ and $z_{u,t}^p$ are common coding message and private coding message, respectively. The streams are allocated corresponding power at the time slot $t$ through $\mathbf{p}_t=\left[\text{p}_{t}^c,\text{p}_{1,t}^p,\ldots,\text{p}_{u,t}^p,\ldots,\text{p}_{|\mathcal{U}|,t}^p\right]$. Let $P_{{\max}}$ denote the maximum transmit power of the BS, the power constraint can be expressed as 
\begin{equation}
    \text{p}_t^c+\sum_{u=1}^{\left|\mathcal{U}\right|}{\text{p}_{u,t}^p} \leq P_{{\max}}.
\end{equation}

Hence, the transmit signal ultimately at the receiver can be expressed as
\begin{equation}
    \mathbf{Z}_t=\sqrt{\text{p}_t^c}z_t^c+\sum_{u=1}^{\left|\mathcal{U}\right|}{\sqrt{\text{p}_{u,t}^p}z_t^p}.
\end{equation}

Due to the impact of the channel, the received signal of  user $u$ at time slot $t$  is expressed as
\begin{equation}
\hat{z}_{u,t}=\sqrt{\text{p}_t^c}\mathbf{h}_{u,t}^Hz_t^c+\mathbf{h}_{u,t}^H\sum_{k=1}^{\left|\mathcal{U}\right|}{\sqrt{\text{p}_t^p}z_t^p}+n_{u,t},
\end{equation}
where $\mathbf{h}_{u,t}$ denotes the channel matrix of user $u$ at the time slot $t$ and $n_{u,t}$ is the corresponding additive white Gaussian noise (AWGN) following $\mathcal{C}\mathcal{N}\left(0,\sigma_{u,t}^2\right)$.

The receiver first decodes the common semantic message stream by treating all private semantic message streams as interference. The SINR of the common semantic message stream decoded by user $u$ at time slot $t$ is
\begin{equation}
    \text{SINR}_{u,t}^c=\frac{\left|\mathbf{h}_{u,t}^H\right|^2\text{p}_t^c}{\sum_{k=1}^{\left|\mathcal{U}\right|}{\left|\mathbf{h}_{u,t}^H\right|^2\text{p}_{k,t}^p}+\sigma_{u,t}^2}.
\end{equation}

After decoding the common semantic message stream, the receiver uses SIC to subtract it. During the decoding of their private semantic message streams, other users' private messages are considered as interference. The common semantic message stream is disregarded in this process since it has already been decoded. The SINR of private semantic message streams decoded by user $u$ at time slot $t$ can be expressed as
\begin{equation}
    \text{SINR}_{u,t}^p=\frac{\left|\mathbf{h}_{u,t}^H\right|^2\text{p}_{u,t}^p}{\sum_{k \neq u}^{\left|\mathcal{U}\right|}{\left|\mathbf{h}_{u,t}^H\right|^2\text{p}_{k,t}^p}+\sigma_{u,t}^2}.
\end{equation}

According to Shannon's theorem, ideally, the transmission rate of all information cannot exceed the Shannon limit. Let $B$ denote bandwidth. The transmission rate of decoded common semantic message stream $R_{u,t}^c$ and private semantic message stream $R_{u,t}^p$ by user $u$ at time slot $t$ can be expressed as
\begin{align}
   R_{u,t}^c=B\log_{2}\left(1+\text{SINR}_{u,t}^c\right),\\
   R_{u,t}^p=B\log_{2}\left(1+\text{SINR}_{u,t}^p\right).
\end{align}

In order to ensure that all users can successfully decode the semantic codeword sequence, the total common semantic message rate cannot exceed the common semantic message rate of any user, i.e, 
\begin{equation}
  R_{t}^c=\min_{u\in\mathcal{U}}R_{u,t}^c.  
\end{equation}

The restored common semantic message stream $s_{u,t}^c$ contains common semantic sub-messages of $|\mathcal{U}|$ users. To maximize transmission efficiency, the BS will adaptively allocate the common message rate to each user, i.e,
\begin{equation}
    \sum_{u=1}^{|\mathcal{U}|}C_{u,t}=R_{t}^c,
\end{equation}
where $C_{u,t}$ is the portion of the total semantic message rate allocated to user $u$ for $s_{u,t}^c$. Thus, the achievable transmission rate of user $u$ at time slot $t$ is
\begin{equation}
    R_{u,t}=R_{u,t}^p+C_{u,t}.
\end{equation}


All users perform SIC to obtain the received codeword sequence is ${\hat{\mathbf{s}}}_{u,t}$ on the received signals. The recovered frame ${\hat{\mathbf{x}}}_{u,t}$ is reconstructed through Deep JSCC decoding and semantic recovery, which can be expressed as
\begin{equation}
{\hat{\mathbf{y}}}_{u,t}=f_\text{JD}\left({\hat{\mathbf{s}}}_{u,t}\right)\ \text{and} \ {\hat{\mathbf{x}}}_{u,t}=f_\text{SR}\left({\hat{\mathbf{y}}}_{u,t}\middle|{\hat{\mathbf{c}}}_{u,t}\right),\label{eq6}
\end{equation}
where $f_\text{JD}$ and $f_\text{SR}$ denote the function of semantic restorer and Deep JSCC decoder of primary link, respectively.

The APVST network is adept at autonomously learning the inter-frame correlations in panoramic videos, which is crucial for both semantic extraction and recovery. This network is specially designed to consider key evaluation metrics during the transmission of panoramic frames, using a weighted spatial attention map to significantly enhance the quality of the immersive experience. The encoding and decoding processes are meticulously guided by the entropy model and the latitude adaptive module, enabling targeted automatic encoding for specific regions. Additionally, the model employs RSMA for the allocation of common and private messages, achieving efficient transmission of panoramic semantic message streams and further enhancing the system capacity in panoramic video transmission scenarios.

\subsection{Other Enable Modules for APVST}
In the description of the transmission process of the system model, we mention several key modules, including the WA module and latitude adaptive module, which are crucial for the efficient transmission of panoramic videos. In the following, we will provide a detailed explanation of the working principles of these modules and their roles within the system.
\subsubsection{WA module}
To measure the quality of panoramic video in the observation space more accurately, the evaluation metrics WS-PSNR and WS-SSIM take into account the panoramic frames structure information \cite{WS_PSNR}, \cite{WS_SSIM}. These metrics combine the distortion relationship between projection planes and the user's FoV to map the spherical distortion to the 2D plane distortion, which can describe the immersive subjective experience quality more accurately in viewing the panoramic video. The specific definition can be expressed as
\begin{align}
\text{WS-PSNR}&=10\log{\left(\frac{\text{MAX}}{\text{WMSE}}\right)}(\text{dB}), \nonumber \\ 
\text{WS-SSIM}&=-10\log{\left(1-\frac{\sum_{k=0}^K\text{WSSIM}_k}{\sum_{k=0}^K {\rm{w}}_{k}^c}\right)}(\text{dB}) \nonumber,
\end{align}
where $\text{MAX}$ represents the maximum value among all pixel points. $K$ means that the frame is divided into $K$ windows and ${\text{w}}_{k}^c$ is the weight of the center point of $k$-th sliding window. $\text{WMSE}$ and $\text{WSSIM}_k$ can be expressed as
\begin{align}
\text{WMSE}&=\frac{\sum_{i=1}^{H}\sum_{j=1}^{W}\left[\left({\hat{x}}(i,j)-x(i,j)\right)^2\times w(i,j)\right]}{\sum_{i=1}^{H}\sum_{j=1}^{W}w(i,j)}, \nonumber \\
\text{WSSIM}_k&=\left( \frac{2\mu_{\rm{x}_k}\mu_{\hat{\rm{x}}_k} + c_1}{\mu_{\rm{x}_k}^2 + \mu_{\hat{\rm{x}}_k}^2 + c_1} \right) \left( \frac{2\sigma_{{\rm{x}_k}\hat{\rm{x}}_k} + c_2}{\sigma_{\rm{x}_k}^2 + \sigma_{\hat{\rm{x}}_k}^2 + c_2} \right) \times {\rm{w}}_{k}^c, \nonumber
\end{align}
where the frame size is $H\times W$. ${\hat{x}}(i,j)$ and $x(i,j)$ represent the pixel values at coordinates $\left(i,j\right)$ in the reconstructed and the original frames, respectively. $\mu$ and $\sigma$ represent the mean and variance, respectively. $c_1$ and $c_2$ are constants. Besides, the weight values $w_{ij}$ decrease gradually from the equator to the poles, and the distribution of $w_{ij}$ is given as 
\begin{equation}
w(i,j)=\cos{\left(\left(i-\frac{H}{2}+\frac{1}{2}\right)\times\frac{\pi}{H}\right)}.\label{eq9}
\end{equation}

Inspired by the spatial attention module which is proposed in \cite{CBAM}, to obtain higher WS-PSNR and WS-SSIM which mean the immersive experience of users, we combine initial weight map $\mathbf{w}$ and feature map $\mathbf{y}_t$ to achieve weighted spatial attention by the neural networks. The weighted spatial attention map $\mathbf{M}_t$ in the primary link can be obtained as
\begin{align}
\mathbf{M}_{u,t} &\!=\!\sigma\!\left(\text{Conv}_{3\times3}^2\left\{\text{AvgPool}^4\left(\mathbf{w}\!\right);\text{MaxPool}^4\left(\mathbf{w}\right);\mathbf{y}_{u,t}\right\}\right) \nonumber \\
&\!=\!\sigma\left(\text{Conv}_{3\times3}^2\left\{\mathbf{F}_\text{Avg};\mathbf{F}_\text{Max};\mathbf{y}_{u,t}\right\}\right),\label{eq8}
\end{align}
where $\text{AvgPool}^4$ and $\text{MaxPool}^4$ denote that the initial weight maps $\mathbf{w}=\{w(i,j)\}\in\mathbb{R}^{H \times W}$ is downsampled by successive average pooling and maximum pooling four times. $H$ and $W$ represent the height and width of panoramic frames, respectively. Besides, a dimension is inserted into the pooling results to represent the number of channels, so the generated feature weight distributions are $\mathbf{F}_\text{Avg}\in\mathbb{R}^{1\times\frac{H}{16}\times\frac{W}{16}}$ and $\mathbf{F}_\text{Max}\in\mathbb{R}^{1\times\frac{H}{16}\times\frac{W}{16}}$, respectively. $\text{Conv}_{3\times3}^2$ denotes two successive convolution operations with convolution kernel of $3\times3$, and $\sigma$ is the Sigmoid activation function. 

Similar to the primary link, the weighted spatial attention map $\mathbf{M}_t^\text{ml}$ in motion link can be obtained as
\begin{equation}
\begin{aligned}
\mathbf{M}_{u,t}^\text{ml} &\!=\!\sigma\left(\text{Conv}^{3\times3}\left\{\text{AvgPool}^4\left(\mathbf{w}\right)\!;\!\text{MaxPool}^4\left(\mathbf{w}\right)\!;\!\mathbf{y}_{u,t}^\text{ml}\right\}\right)\\
&\!=\!\sigma\left(\text{Conv}^{3\times3}\left\{\mathbf{F}_\text{Avg};\mathbf{F}_\text{Max};\mathbf{y}_{u,t}^\text{ml}\right\}\right).\label{eq11}
\end{aligned}
\end{equation}

The feature information after deflation is obtained by multiplying the feature map with the points corresponding to the weighted spatial attention map. This process can be described as ${\mathbf{y^\prime}}_{u,t}^\text{ml}=\mathbf{y}_{u,t}^\text{ml}\otimes\mathbf{M}_{u,t}^\text{ml}$ in the motion link and $\mathbf{y}_{u,t}^\prime=\mathbf{y}_{u,t}\otimes\mathbf{M}_{u,t}$ in the primary link, respectively.

By utilizing the potent learning capabilities of neural networks, a correlation is established between weight maps and feature maps. Similar to the attention mechanism, this process assigns weights to each point on the feature map and enhances attention to regions with higher weight values during the information recovery phase. Consequently, such an approach leads the quality of the user's immersive experience WS-PSNR and WS-SSIM to be further improved.

\subsubsection{Latitude Adaptive Module}
Due to the characteristics of ERP, it stretches the pixels at different latitudes to different degrees. According to the work in \cite{End_to_End_Optimized_360_Image_Compression}, the higher latitude needs more pixels to represent it. In other words, it needs more bits to represent high-latitude pixels if a uniform compression strategy is used. 

To reduce the information redundancy of transmission and improve the transmission efficiency of panoramic video, we propose the latitude adaptive network. For such a network, we design the adaptive weight feature map $\boldsymbol{\omega}_{u,t}=\left\{\omega_{u,t}(m,n)\right\}\in\mathbb{R}^{\frac{H}{16}\times\frac{W}{16}}$, which can be expressed as
\begin{equation}
\omega_{u,t}(m,n)=\eta_{u,t}(m,n){w}_\text{AAP}(m,n)+1-\eta_{u,t}(m,n),\label{eq16}
\end{equation}
where $\omega_{u,t}(m,n)$ denotes the adaptive weight feature value corresponding to the feature point $\left(m,n\right)$ at time $t$. ${w}_\text{AAP}(m,n)$ denotes the weight value of $\mathbf{w}$ after the adaptive average pooling (AAP) operation. $\eta_{u,t}(m,n)$ denotes the adaptive weight factor value of weight factor map $\boldsymbol{\eta}_{u,t}=\left\{\eta_{u,t}(m,n)\right\}\in\mathbb{R}^{\frac{H}{16}\times\frac{W}{16}}$, which is learned by the network according to the magnitude of entropy information. It can be expressed as $\boldsymbol{\eta}_{u,t}=f_\text{AW}\left(\mathbf{e}_{u,t}\right)$, where $f_\text{AW}$ denotes the function of adaptive weight factor network and $\mathbf{e}_{u,t}=\left\{e_{u,t}(m,n)\right\}\in\mathbb{R}^{\frac{H}{16}\times\frac{W}{16}}$ means the entropy map corresponding to the feature map. 

In order to achieve the adaption of feature information dimension in latitude, the dimensions of the information in the wireless channel cannot exceed the limit of the maximum dimension given in the corresponding latitude. This constraint can be expressed as
\begin{equation}
l_{u,t}(m,n)\le \omega_{u,t}(m,n)\times {\max}\left(\mathcal{Q}\right),\label{eq17}
\end{equation}
where $l_{u,t}(m,n)$ denotes the information dimension corresponding to feature point $\left(m,n\right)$ by quantizing, which can be expressed as the number of channels. ${\max}\left(\mathcal{Q}\right)$ is the maximum value of the quantized set, which is given in Section \ref{section experiment}.
Therefore, the loss function of this module is expressed as
\begin{equation}
\mathcal{L}_{u,t}^\text{la}=\sum_{m}\sum_{n}{{\max}\left(0,\omega_{u,t}(m,n)\times {\max}\left(\mathcal{Q}\right)-l_{u,t}(m,n)\right)}.\label{eq18} 
\end{equation}

However, the quantization operation tends to cause gradient dispersion \cite{End_to_end_optimized_image_compression}, we convert the dimensional restriction to the entropy restriction in the training phase, i.e,
\begin{equation}
\mathcal{L}^\prime{_{u,t}^\text{la}}\!=\!\sum_{m}\sum_{n}{{\max}\left(0,\omega_{u,t}(m,n)\!\times\!{\max}\left(\mathbf{e}_{u,t}\right)\!-\!e_{u,t}(m,n)\right)}.\label{eq19} 
\end{equation}

With the latitude adaptive module, the entropy model can more accurately compute the entropy of feature maps in different latitudes and achieve efficient transmission of panoramic videos.
\subsection{Problem Formulation}
For the above description of the system model, we decouple it into two problems and optimize them using different objective functions and methods. The optimization goal for the semantic transmission network APVST is to maximize users' immersive experience quality with minimal bandwidth consumption. Meanwhile, the optimization goal for RSMA is to maximize all users' QoS scores based on semantic communication.

\subsubsection{The optimization goal of APVST}
The APVST network aims to maximize the transmission quality of frames with the minimum channel bandwidth overhead. In fact, the frames are transmitted continuously, $N$ consecutive frames are one group of pictures (GoP). In the training phase, the previously recovered panoramic frame is used as the reference frame for the current frame to achieve panoramic video transmission. Therefore, the loss function is formulated as 
\begin{align}
\mathcal{L}_u\!=\!\frac{1}{N}\!\sum_{t=1}^{N}\left[{D\left(\mathbf{x}_{u,t}, \hat{\mathbf{x}}_{u,t}\right)\!+\!\alpha\!\left(\mathcal{L}_{u,t}^{\text{enml}}\!+\!\mathcal{L}_{u,t}^{\text{en}}\!\right)\!+\!\beta\mathcal{L}^\prime{_{u,t}^\text{la}}}\right],\label{all loss of apvst}
\end{align}
where $D\left(\mathbf{x}_{u,t}, \hat{\mathbf{x}}_{u,t}\right)$ represents the magnitude of image distortion, which is chosen as the WMSE for WS-PSNR, and as the inverse of WS-SSIM for WS-SSIM to indicate the distortion of frames \cite{WS_PSNR}, \cite{WS_SSIM}. To minimize channel bandwidth overhead, we take the entropy $e_{u,t}(m,n)$ and $e_{u,t}^\text{ml}(m,n)$ into account. $\mathcal{L}_{u,t}^{\text{enml}}=\sum_{m}\sum_{n}{e_{u,t}^\text{ml}(m,n)}$ and $\mathcal{L}_{u,t}^{\text{en}}=\sum_{m}\sum_{n}{e_{u,t}(m,n)}$ represent the channel bandwidth cost of motion link and primary link, respectively. $\alpha$ and $\beta$ denote the balance coefficients, which are constants. In section \ref{internal structure of apvst}, the internal structure of the APVST network will be introduced, and the optimization of the network is completed based on this loss function.

\subsubsection{Problem Formulation of RSMA-enabled semantic information transmission}
In the process of panoramic video transmission, the factors that most significantly affect the QoS are the transmission delay and the quality of the immersive experience for the users. To comprehensively consider these two components, QoS is transformed into a scoring model which is decomposed into two parts: the transmission delay score and the immersive experience quality score.

To better represent the data size of the semantic codeword sequence to be transmitted at time slot $t$, referring to \cite{data_size}, let $\lambda$ denote the number of bytes per semantic codeword. The data size for user $u$ at time slot $t$ is given as
\begin{equation}
    D_{u,t}=\lambda k_{u,t},
\end{equation}
where $k_{u,t}$ represents the dimension of the semantic codeword sequence $\mathbf{s}_{u,t}$ for user $u$ at time slot $t$, which can be adjusted by selecting different values from the quantization set $\mathcal{Q}$. Generally speaking, the quality of the user's immersive experience increases with the increase of channel bandwidth ratio (CBR) \cite{deepsc}, which can be expressed as $\frac{k_{u,t}}{H\times W\times3}$, where 3 represents the original panoramic frame has three RGB channels. The transmission delay of user $u$ at time slot $t$ can be expressed as
\begin{equation}
    T_{u,t}=\frac{D_{u,t}}{R_{u,t}}=\frac{\lambda k_{u,t}}{R_{u,t}^p+C_{u,t}}.
\end{equation}

Transmission delay plays a key role in users' QoS, yet it exhibits two extreme cases. On one hand, when latency is minimal, further reductions in delay do not significantly enhance QoS. On the other hand, excessive delays become intolerable, leading to a substantial degradation in QoS to unacceptable levels. In order to approximate this trend and jointly optimize it with the immersive experience quality below, we refer to \cite{latency_to_score} and convert it into a scoring model. The transmission delay score of user $u$ at time slot $t$ can be expressed as
\begin{equation}
    F_{u,t}^T=\begin{cases} 
  \frac{1}{1 + e^{-\zeta (T_{{\max}} - T_{u,t})}}, &T_{u,t} \leq T_{\max} \\
  0, & T_{u,t} \geq T_{\max} 
\end{cases},
\label{latency to score}
\end{equation}
where $T_{{\max}}$ is the the maximized tolerable transmission delay.  $\zeta$ is a constant deciding the steepness of the satisfactory curve.

The quality of the immersive experience is also one of the key factors affecting users' QoS. In the considered scenario of panoramic video transmission, the immersive experience quality $Q_{u,t}$ is related to the semantically encoded CBR and the user's SNR, i.e,
\begin{equation}
    Q_{u,t}=f_{\text{APVST}}\left(o_{u,t},\vartheta_{u,t}\right)
\end{equation}
where $f_{\text{APVST}}$ means the function of the APVST network.  $Q_{u,t}$ is chosen as WS-PSNR or WS-SSIM.  $o_{u,t}$ and $\vartheta_{u,t}$ are CBR and SNR for user $u$ at time slot $t$, respectively. To accelerate the optimization of the problem, we employ polynomial regression, nonlinear regression, and random forests to approximate the fitting of $f_{\text{APVST}}$.

Similar to transmission delay, there are two extremes in immersive experience quality. In order to perform joint optimization with the above transmission delay, it is converted into a score through maximum-minimum quantization. The immersive experience quality score $S_{u,t}^Q$ for user $u$ at time slot $t$ can be expressed as
\begin{equation}
    F_{u,t}^Q=\begin{cases} 
  1, & Q_{u,t} \geq Q_{\max} \\
  \frac{Q_{u,t}-Q_\text{min}}{Q_{\max}-Q_\text{min}}, & Q_\text{min} < Q_{u,t} < Q_{\max} \\
  0, & Q_{u,t} \leq Q_\text{min} 
  \end{cases},
\end{equation}
where $Q_\text{min}$ and $Q_{\max}$ are the upper and lower limits of immersive experience quality, respectively.
Then the RSMA-enabled semantic information transmission optimization problem can be expressed as
\begin{subequations}
\renewcommand{\theequation}{30} 
\begin{align}
\max_{\mathbf{p}_t, \mathbf{o}_t, \mathbf{C}_t} \quad & \sum_{u=1}^{\left|\mathcal{U}\right|} F_{u,t} = \sum_{u=1}^{\left|\mathcal{U}\right|} \left[\kappa F_{u,t}^T+(2-\kappa)F_{u,t}^Q\right] \tag{30} \label{max problem}
\end{align}
\vspace{-2mm}
\renewcommand{\theequation}{30\alph{equation}} 
\begin{align}
\text{s.t.} \quad & p_t^c+\sum_{u=1}^{\left|\mathcal{U}\right|}p_{u,t}^p \leq P_{{\max}}, \label{lima}\\
& 0<o_{u,t} \leq o_{\max}, \label{limb} \\
& \sum_{u=1}^{\left|\mathcal{U}\right|}C_{u,t}=R_{t}^c, \label{limc}
\end{align}
\end{subequations}
where $\mathbf{o}_t=\left[o_{1,t},\ldots,o_{u,t},\ldots,o_{\left|\mathcal{U}\right|,t}\right]$ is the CBR verctor at time slot $t$, and $\mathbf{C}_t=\left[C_{1,t},\ldots,C_{u,t},\ldots,C_{\left|\mathcal{U}\right|,t}\right]$ is the common rate allocation vector at time slot $t$. $\kappa$ is the balance coefficient between the transmission delay score and the immersive experience quality score. In the above optimization problem, constraint \eqref{lima} means that the power of the precoding matrix at time slot $t$ can't exceed the maximum transmit power. Constraint \eqref{limb} means the CBR can't exceed the maximum of CBR $o_{\max}$, where $o_{\max}$ is decided by quantized set $\mathcal{Q}$. \eqref{limc} is common rate allocation constraint to ensure that the common message stream can be successfully decoded by all users. 

We can see that problem \eqref{max problem} is a mixed continuous-discrete non-convex optimization problem. Additionally, the time-varying nature of the channel and the variations in the user's FoV further complicate the resolution of the problem. The solutions proposed by \cite{vr_meet_rsma} are not applicable to our considered scenario. This is due to the fact that these approaches do not take into account the temporal impacts on the environment, nor do they optimize the transmission of information at a semantic level. Direct application of these solutions to our scenario may lead to local optima, thereby degrading the system's transmission performance and affecting the users' QoS. Considering the continuity of time, and since the transmitted information is video, there exists a certain semantic correlation between consecutive frames. Moreover, the user's FoV continuously moves over time. Based on these analyses, we can transform the problem into a MDP. By making appropriate decisions at different time instances, we aim to maximize the total user score. Furthermore, we propose a solution based on PPO-based algorithm, the specifics of which will be detailed in Section \ref{DRL section}.

\begin{figure*}[htbp]
    \centering
        \subfloat[Network structure of semantic transmitter]{\includegraphics[width=83mm]{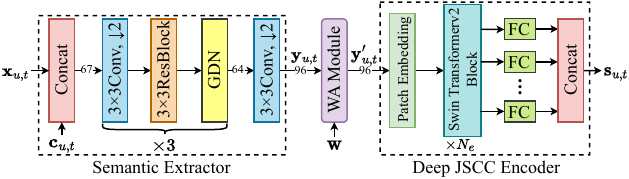}}%
    \hfill
        \subfloat[Network structure of semantic receiver]{\includegraphics[width=92mm]{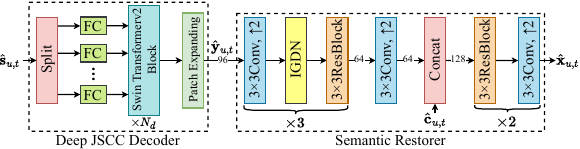}}
    \caption{Network structures of the primary link in APVST. $k\times k$ Conv is a convolution with $k\times k$ filters, and the output channels of convolution are given on horizontal line. $\uparrow2$ and $\downarrow2$ indicate upsampling and downsampling with a stride of 2. GDN denotes the Generalised Divisive Normalization in \cite{density_modeling_images}, IGDN denotes the inverse operation of GDN.}
    \label{network of system}
\end{figure*}

\section{Internal Structure and Training Procedure of APSVT} \label{internal structure of apvst}
In this section, we introduce the internal structure and training procedure of the APVST network. The disassembled parts of the internal structure are shown in Fig.~\ref{network of system} and Fig.~\ref{network of other modules}. Since the structures of primary and motion links are similar, the primary link will be introduced in detail.

\subsection{The Overall Structure of APSVT}
Referring to \cite{DCVC}, the current panoramic frame $\mathbf{x}_t$ is firstly concatenated with the context $\mathbf{c}_t$. Then, we achieve semantic extraction by successive downsampling convolutions, residual structure (Resblock), and GDN. In semantic recovery, as opposed to semantic extraction, the recovered feature map is upsampled successively by convolution with IGDN and Resblocks. Finally, we fuse ${\hat{\mathbf{c}}}_t$ with the upsampling result and adjust the number of channels by convolution layer and Resblocks.

In particular, the scale-space motion estimation in motion link refers to the scale-space flow, which is proposed in \cite{Scale_space_flow}. Compared to the traditional optical flow, it takes into account the scale space to generate scale parameters and act on the scale-space warp. Consequently, the network can achieve motion estimation efficiently when performing complex scenarios involving occlusion and rapid object movement.

The Deep JSCC structure employs symmetric encoding and decoding techniques. Drawing inspiration from \cite{DVST}, and to further augment the capacity of Deep JSCC in capturing long-term correlations, we integrate the swin transformer v2 as the backbone of both the Deep JSCC encoder and decoder, and take multi-head self-attention (MHSA) as the core of the network. In contrast to the original swin transformer, the swin transformer v2 utilizes cosine similarity and nonlinear relative position bias to enhance the network's robustness against various downstream tasks \cite{swin_transformerv2}.

The Deep JSCC encoder is guided by the entropy model and the latitude adaptive module to achieve the variable-length encoding of the feature map. The information dimension $l_{t,mn}$ is derived by quantizing the entropy $e_{t,mn}$ into the set of quantized values, which can be expressed as $l_{t,mn}\!\in\!\mathcal{Q}\!=\!\left\{q_1,q_2,\cdots,q_I\right\}$ with size $I$. The feature information dimension is adjusted by multiple learnable fully connected (FC) layers, and power normalization is performed before the information is transmitted into the channel.

\subsection{The Structure of Enhanced Modules}
\begin{figure}[t]
    \centering
        \subfloat[]{\includegraphics[width=22mm]{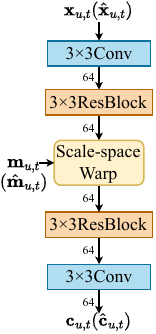} \label{network of cg}} 
    \hfill
        \subfloat[]{\includegraphics[width=34mm]{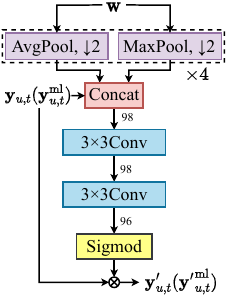} \label{network of wa}} 
    \hfill
        \subfloat[]{\includegraphics[width=24mm]{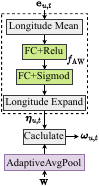} \label{network of la}} 
    \caption{Detailed structures of (a) Context Generator, (b) WA Module, and (c) Latitude Adaptive Module.}
    \label{network of other modules}
\end{figure}

Fig.~\ref{network of other modules} shows in detail the structure of modules which are used to enhance the panoramic video transmission efficiency.

The context generation module generates context $\mathbf{c}_t$ and ${\hat{\mathbf{c}}}_t$ based on motion information and reference panoramic frame. This ensures that APVST can perform long-term learning and stable prediction. The structure of this network module is depicted in Fig.~\ref{network of cg}. Since the reference frame is subjected to convolution and Resblock, a scale-space warp operation is applied to capture the complex motion information. To address the issue of semantic information discontinuity caused by the warp operation and effectively generate the context, the warped result is further refined using Resblock and convolution.

The proposed WA module deflates the feature maps appropriately to enhance the final video quality evaluation metrics, which structure is illustrated in Fig~\ref{network of wa}. To maximize the retention of weight map information, we combine the original weight map $\mathbf{w}$ and the semantic feature map $\mathbf{y}_t$. Due to the different sizes of $\mathbf{w}$ and $\mathbf{y}_t$, we employ two different pooling downsampling methods (four consecutive average pooling and maximum pooling) to act on $\mathbf{w}$. The obtained result is concatenated with $\mathbf{y}_t$, and the weighted spatial attention map $\mathbf{M}_t$ is finally obtained by the Sigmoid activation function after two layers of convolution operation. The first convolution layer is designed to learn the relationship between $\mathbf{M}_t$ and $\mathbf{y}_t$, while the second layer adjusts the number of channels.

The application of the latitude adaptive module enhances the transmission efficiency of panoramic videos. By utilizing function $f_{\text{AW}}$, the adaptive weight feature map $\boldsymbol{\omega}_t$ is derived from entropy map $e_{t,mn}$ and original weight map $\mathbf{w}$. The detailed module structure is shown in Fig~\ref{network of la}.  Given that the semantic feature map information is more latitude-sensitive, a conversion from $\mathbf{e}_t\in\mathbb{R}^{\frac{H}{\mathbf{16}}\times\frac{W}{\mathbf{16}}}$ to $\mathbb{R}^{\frac{H}{\mathbf{16}}}$ is performed by averaging over longitude. Subsequently, the weight factor map $\boldsymbol{\eta_t}$ is generated through FC layers using ReLU and Sigmoid activation functions and an expansion along the longitude is applied to transform $\mathbb{R}^{\frac{H}{\mathbf{16}}}$ back to $\mathbb{R}^{\frac{H}{\mathbf{16}}\times\frac{W}{\mathbf{16}}}$. Finally, $\boldsymbol{\eta_t}$ and $\mathbf{w}$ are merged to compute $\boldsymbol{\omega}_t$ according to equation \eqref{eq16}.

\subsection{Training Procedure}
Referencing \cite{DVST}, considering the complexity of the APVST network architecture and the difficulty in training transformer networks, we employ a block-wise training approach to optimize the performance of APSVT. The ``$\sim$'' mentioned below represents that the information has only undergone semantic extraction and restoration (with noise added) without being the Deep JSCC encoding and decoding. Overall, the training procedure for the APVST network can be described as follows:
\begin{enumerate}[label=(\arabic*),leftmargin=*]
    \item Based on the pretrained network for scale-space flow estimation, we initially train the networks for $f^\text{ml}_\text{SE}$, $f^\text{ml}_\text{SR}$, and the entropy model of the motion link. The loss function for this step can be expressed as
    \begin{equation}
        {\mathcal{L}^\prime}^\text{ml}_{u,t}={D^\prime}^\text{ml}_{u,t}+\alpha\mathcal{L}_{u,t}^{\text{enml}},
    \end{equation}
    where ssw denotes the scale space warp function in context generator and ${D^\prime}^\text{ml}_{u,t}=D\left(\mathbf{x}_{u,t}, \text{ssw}\left(\mathbf{x}_{u,t-1},\tilde{\mathbf{m}}_{u, t}\right)\right)$ represents the reconstruction loss of the training network.
    
    \item Continuing to consider the encoding and decoding networks of Deep JSCC, we will proceed with training the $f^\text{ml}_\text{JE}$, $f^\text{ml}_\text{JD}$, rate allocation, and WA module networks within the motion link. To ensure the stability of training, we will incorporate the reconstruction loss from step (1). Overall, the loss function for this step is as follows:
    \begin{equation}
        \mathcal{L}^\text{ml}_{u,t}=D^\text{ml}_{u,t}+{D^\prime}^\text{ml}_{u,t}+\alpha\mathcal{L}_{u,t}^{\text{enml}},
    \end{equation}
    where $D^\text{ml}_{u,t}=D\left(\mathbf{x}_{u,t}, \text{ssw}\left(\mathbf{x}_{u,t-1},\hat{\mathbf{m}}_{u, t}\right)\right)$ represents the reconstruction loss of the motion link.

    \item Upon completing the training of the motion link networks, we will freeze these network parameters. Simultaneously, training will commence for the $f_\text{SE}$, $f_\text{SR}$, and entropy model networks in the primary link. Similar to the loss function in step (1), the loss function for this step can be expressed as
    \begin{equation}
        \mathcal{L}^\prime_{u,t}=D^\prime_{u,t}+\alpha\mathcal{L}_{u,t}^{\text{en}},
    \end{equation}
    where $D^\prime_{u,t}=D\left(\mathbf{x}_{u,t}, \tilde{\mathbf{x}}_{u,t}\right)$ represents the reconstruction loss of the training network.

    \item Keeping the above settings unchanged, we will continue to train the $f_\text{JE}$, $f_\text{JE}$, rate allocation, WA module, and latitude adaptive module networks within the primary link. The training loss for this step is formulated as
    \begin{equation}
        \mathcal{L}_{u,t}=D_{u,t}+{D^\prime}_{u,t}+\alpha\mathcal{L}_{u,t}^{\text{en}}+\beta{\mathcal{L}^\prime}^\text{la}_{u,t},
    \end{equation}
    where $D^\prime_{u,t}=D\left(\mathbf{x}_{u,t}, \hat{\mathbf{x}}_{u,t}\right)$ represents the reconstruction loss of the primary link.

    \item Finally, we will unfreeze the parameters of all networks and train the APVST network based on the loss function \eqref{all loss of apvst}, as specified in the optimization objectives.
\end{enumerate}

\section{PPO for RSMA-Enabled transmission of semantic streaming} \label{DRL section}

In this section, we transform the problem of RSMA-enabled panoramic video semantic information transmission into an MDP. We employ PPO-based algorithm to solve for the optimal solution, ensuring that the policy can select appropriate actions under any state to maximize the users' QoS.

\subsection{Markov Decision Process}
In the sequential MDP, the most important is to model four elements, i.e. state $\mathcal{S}_t$, action $\mathcal{A}_t$, reward $r_t$, and transition equation. The MDP can be simply described as follows:
\begin{itemize}
    \item Choose appropriate action $\mathcal{A}_t$ based on current state $\mathcal{S}_t$.
    \item According to the current state $\mathcal{S}_t$ and action $\mathcal{A}_t$, enter the next state  $\mathcal{S}_{t+1}$ according to the transition equation.
    \item Compute the reward $r_t$ based on the action $\mathcal{A}_t$ and the current state $\mathcal{S}_t$.
    \item Adjust policy based on rewards to maximize rewards.
\end{itemize}

The goal of the MDP is to maximize long-term rewards and to make appropriate decisions under changing conditions. Based on this, we transform problem \eqref{max problem} into a MDP. The state vector space $\mathcal{S}_t$ can be expressed as
\begin{equation}
\mathcal{S}_t=\left[\mathbf{H}_t,\mathcal{I}_t^s,\mathbf{p}_t,\mathbf{o}_t,\mathbf{C}_t,\mathbf{R}_t^p\right],
\end{equation}
where $\mathbf{H}_t=\left[\left|\mathbf{h}_{1,t}^H\right|^2,\ldots,\left|\mathbf{h}_{u,t}^H\right|^2,\ldots,\left|\mathbf{h}_{\left|\mathcal{U}\right|,t}^H\right|^2\right]$ is the channel vector for all users. $\mathcal{I}_t^s=\left[\mathbf{I}_{1,t}^s,\ldots,\mathbf{I}_{u,t}^s,\ldots,\mathbf{I}_{\left|\mathcal{U}\right|,t}^s\right]$ denotes the FoV information on semantic encoding level for all users. $\mathbf{R}_t^p=\left[R_{1,t}^p,\ldots,R_{u,t}^p,\ldots,R_{\left|\mathcal{U}\right|,t}^p\right]$ is the private message rate of users. The initialization of MDP will be implemented corresponds to time $t=0$. By including as much environment and action information as possible in the state, the policy model can choose the appropriate resource allocation action. Therefore, the action vector space can be defined as
\begin{equation}
\mathcal{A}_t=\left[\mathbf{p}_t,\mathbf{o}_t,\mathbf{C}_t\right],
\end{equation}
where we consider three types of information that together constitute the action space vector. $\mathbf{p}_t$ corresponds to the selection of different transmit powers for the common and private semantic message streams at time slot $t$, based on the state vector space. $\mathbf{o}_t$ represents the choice of an appropriate CBR. If the CBR is set too high, it leads to increased data transmission without a corresponding gain in users' immersive experience quality. Conversely, a CBR with a too small value results in a drastic decline in the quality of the immersive experience. $\mathbf{C}_t$ signifies the allocation of appropriate common transmission rates for each user, guided by users' semantic-level FoV information.

Examining the objective function that needs to be maximized in problem \eqref{max problem}, it can be characterized as finding the optimal solution of a piecewise function under certain constraints. In terms of designing the reward function, we aim to maximize the average total score of users. Consequently, the reward function at time slot t can be expressed as
\begin{equation}
    r_t\left(\mathcal{S}_t, \mathcal{A}_t\right)=\frac{1}{\left|\mathcal{U}\right|}\sum_{u=1}^{\left|\mathcal{U}\right|} F_{u,t} =\frac{1}{\left|\mathcal{U}\right|}\sum_{u=1}^{\left|\mathcal{U}\right|} \left[\kappa F_{u,t}^T+(2-\kappa)F_{u,t}^Q\right].
\end{equation}

In the aforementioned reward function, since both the transmission delay score and the immersive experience quality score are normalized, the maximum value of the reward function is 2. In practical MDP, variations in the reward function will guide the policy model to make corresponding actions in different states, thereby achieving the maximum total scores of users. Moreover, for different scenarios, the balance coefficient $\kappa$ can be adjusted to suit various environments. For instance, in scenarios with high delay sensitivity, $\kappa$ can be increased to prioritize optimizing the user's transmission delay score. For scenarios where the quality of immersive experience is more critical, $\kappa$ can be decreased to focus on optimizing the user's immersive experience quality. Overall, the design of the reward function is flexible enough to adapt to different scenarios, offering a degree of universality.

\begin{figure}[bp]
\centering
\includegraphics[width=85mm]{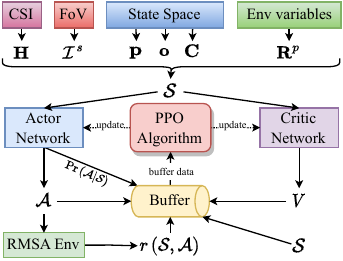}
\caption{The training framework of PPO-based approach.}
\label{DRL}
\end{figure}
\subsection{PPO for MDP Formulation}
Based on the MDP formulation, we adopt a PPO-based algorithm to address the MDP problem. The training framework of PPO-based approach is shown in Fig.~\ref{DRL}. Let $l$ denote $l$-th update. There are two trainable networks, which include actor network $\pi_{\theta^a_l}$ and critic network $\phi_{\theta^c_l}$, where $\theta^a_l$ and $\theta^c_l$ are the trainable parameters of the actor network and the critic network, respectively. The actor represents the policy network in the MDP, which will provide corresponding actions based on the current state $\mathcal{S}_t$ during training. Meanwhile, the critic assesses the current state $\mathcal{S}_t$ and action $\mathcal{A}_t$ to produce a judgment value.

The state space in PPO is similar to the state information space in MDP, i.e. $\mathcal{S}_t$. According to the \cite{channel_gain}, in terms of channel gain, we consider both path loss and Rayleigh fading. $\left|\mathcal{U}\right|$ users will be randomly scattered within a circular area with radius $R$, with each user's scattering following an independent random Poisson process. Therefore, the channel gain can be described as
\begin{equation}
    \mathbf{h}_{u,t} = \mathbf{h}^\prime_{u,t} \times 10^{-\frac{PL_{u,t}}{20}}, 
\end{equation}
where $\mathbf{h}^\prime_{u,t}$ represents the channel gain produced by Rayleigh fading for user $u$ at time slot $t$, which can also represent small-scale fading during the transmission process. $PL_{u,t}=32.4+20\log_{10}f_c+21\log_{10}d_{u,t}$ represents the path loss generated by user $u$ at time slot $t$, which can also represent large-scale fading during the signal transmission process. $d_{u,t}$ and $f_c$ represent the distance between the user and the BS, and the carrier frequency of the user channel, respectively.

\begin{algorithm}[t]\label{training pipeline}
    \caption{PPO-based Training Pipeline}
    Initialize the trainable parameters of $\theta^a_0$, $\theta^c_0$. \\
    Set global timestep $g=0$, the timesteps for parameter update $\delta$, and the update times $l=0$.\\
    \For{$k=0,1,2,\ldots$}{
        Randomly generate the channel gains $\mathbf{H}_t, t \in [kT,$ $(k+1)T)$ during transmission. \\
        Retrieve the user's FoV information and compute the corresponding semantic-level FoV information $\mathcal{I}_t^s, t \in \left[kT, (k+1)T\right)$. \\
        Initialize the environment to obtain the initial state $\mathcal{S}_{kT}$. \\
        \For{$t=kT, kT+1, kT+2, \ldots, (k+1)T-1$}{
            Make action decisions using the actor network, denoted as $\mathcal{A}_t= \pi_{\theta^a_l}\left(\mathcal{S}_t\right)$. \\
            Obtain the corresponding action probabilities $\text{Pr}\left(\mathcal{A}_t|\mathcal{S}_t;\theta^a_l\right)$. \\
            Compute the state-action value using the critic network, i.e., $V_t=\pi_{\theta^c_l}\left(\mathcal{S}_t, \mathcal{A}_t\right)$. \\
            Based on the current state $\mathcal{S}_t$ and action $\mathcal{A}_t$, observe the reward $r_t\left(\mathcal{S}_t, \mathcal{A}_t\right)$. \\
            Obtain the next state $\mathcal{S}_{t+1}$ and the termination flag $D_t$. \\
            Store the tuple $\left(\mathcal{S}_t, \mathcal{A}_t, \text{Pr}\left(\mathcal{A}_t|\mathcal{S}_t;\theta^a_l\right),V_t, r_t, D_t\right)$ into the buffer.\\
            $g\leftarrow g+1$.\\
            
            \If{$g \% \delta$ is equal $0$}{
                Update the actor and the critic network parameters based on Algorithm \ref{ppo update}.\\
            }
        }
    }    
\end{algorithm}

Incorporating the semantic-level FoV information $\mathbf{I}_{u,t}^s$ of users, we treat it as a state for the optimization of the problem. For instance, for a panoramic frame of size $H\times W$, it is divided into a $3\times 3$ grid. Let $\mathbf{J}_{m \times n}$ denote an $m \times n$ matrix where every element is 1. Assuming the current user's FoV is located at the center of the panoramic frame, the FoV information can be expressed as
\begin{equation}
\mathbf{I}_{u,t} = \begin{bmatrix}
    \xi & \xi & \xi \\
    \xi & 1 & \xi \\
    \xi & \xi & \xi
\end{bmatrix} \otimes \mathbf{J}_{\frac{H}{3} \times \frac{W}{3}},
\end{equation}
where the Kronecker product $\otimes$ s used to replicate the matrix $\mathbf{J}_{\frac{H}{3} \times \frac{W}{3}}$ within a larger matrix, following the configuration dictated by the first matrix. In this configuration, the central element remains unscaled by $\xi$, while the surrounding elements are scaled accordingly.

By downsampling FoV information $\mathbf{I}_{u,t}$ through four consecutive average pooling operations with a kernel size of 1 and a stride of 2, we obtain semantic-level FoV information $\mathbf{I}_{u,t}^s$, i.e,
\begin{equation}
\mathbf{I}_{u,t}^s = \begin{bmatrix}
    \xi & \xi & \xi \\
    \xi & 1 & \xi \\
    \xi & \xi & \xi
\end{bmatrix} \otimes \mathbf{J}_{\frac{H}{48} \times \frac{W}{48}},
\end{equation}
where the numerical values may vary due to adjustments in the kernel size, stride, and $\xi$ in actual simulations.

To achieve an optimal decision-making process, PPO focuses on maximizing cumulative long-term rewards. This objective extends beyond short-term gains, aiming for the highest returns over a temporal sequence. The essence of the approach lies in evaluating the overall value brought by a series of state and action sequences. The target reward to be optimized can be expressed as
\begin{equation}
    {\max}_{\theta^a_l}\mathbb{E}\left[\frac{1}{T}\sum_{t=kT}^{(k+1)T-1} r_t\left(\mathcal{S}_t, \mathcal{A}_t\right)\right],
\end{equation}
where $\mathbb{E}(\cdot)$ is the expectation operator, indicating the average reward that is expected to be obtained over time. Let $T$ denote the time span. This time-series-based mechanism allows us to receive immediate feedback on each decision made through interaction with the environment, thereby optimizing the transmission of panoramic video frames. This method is particularly well-suited for optimizing panoramic video transmission problems. The transmission strategy can be automatically adjusted under dynamically changing environmental conditions to ensure the quality of each panoramic frame transmitted. Through continuous learning and adjustment, PPO has the potential to enhance user experience and system performance during the panoramic video transmission process.


Next, the PPO-based algorithm is exploited to address the optimization problem. The training pipeline is as delineated in Algorithm \ref{training pipeline}. Let a span of $T$ time denote one episode. At the beginning of each episode, channel gains are generated and the semantic-level FoV information for users is computed. For environmental initialization, equal power and a common rate are allocated for all message streams and users, yielding the initial state $\mathcal{S}_{kT}$. Subsequently, action $\mathcal{A}_t$, action probability $\text{Pr}\left(\mathcal{A}_t|\mathcal{S}_t;\theta^a_l\right)$, state-action value $V_t$, reward $r_t$, the next state $\mathcal{S}_{t+1}$, and the termination flag $D_t$ can be obtain, where $D_t \in \{0,1\}$ indicates continuation ($0$) or termination ($1$) of the current episode. The acquired values are then stored in the buffer for the subsequent update of network parameters.

The crux of the PPO-based update process is the iterative optimization of the policy network's parameters over multiple iterations. The specific process details are presented in Algorithm \ref{ppo update}. The process begins by extracting the data stored in the buffer as input. To simplify notation, we use $i=0$ to denote the start time of the parameter update within the algorithm. The initial discount reward is set to zero, and cumulative discounted rewards are calculated at each time step. The advantage is determined by the discrepancy between the predicted action-state value and the actual returns, providing direction for the actor network's parameter updates. 

In each iteration, the algorithm first computes the action probabilities and action-state values for the current update time, where $\mathbb{A}$ and $\mathbb{S}$ represent the concatenation of states and actions in the temporal dimension from the buffer, respectively. Based on these parameters, we can define the objective function for the actor network that needs to be optimized as follows
\begin{equation}
    L\left(\theta^a_l,\theta^a_{l_o}\right)=\min\left(u_i, \text{clip}\left(u_i, 1-\epsilon, 1
    +\epsilon\right)\right)\hat{A}_i, \label{l_theta}
\end{equation}
where $u_i=\frac{\text{Pr}\left(\mathcal{A}_i|\mathcal{S}_i;\theta^a_l\right)}{\text{Pr}\left(\mathcal{A}_i|\mathcal{S}_i;\theta^a_{l_o}\right)}$ represents the ratio of action probabilities. The $\text{clip}$ signifies the clipping function, which clips this probability ratio to remain within the range $\left[1-\epsilon, 1+\epsilon\right]$. This clipping is vital for ensuring the stability of the PPO training process. For the critic network, the parameters are updated using a mean squared error loss to bring the evaluation values closer to the cumulative discounted rewards.

\begin{algorithm}[t]  \label{ppo update}
    \caption{PPO-based Update Process}
    \KwInput{The tuple $\left(\mathcal{S}_i, \mathcal{A}_i, \text{Pr}\left(\mathcal{A}_i|\mathcal{S}_i;\theta^a_l\right),V_i, r_i, D_i\right)$ stored in the buffer and update times $l$.}
    Set discount factor $\gamma$, clipping coefficient $\epsilon$, update iterations $J$, and initial update times $l_o=l$. \\
    Set the initial discount reward $G_{\delta}=0$, and compute the cumulative discount rewards at other times, i.e.,
    \begin{equation}
        G_i = r_i+\gamma (1-d_i)G_{i+1}, i=\delta-1, \delta-2, \ldots, 1, 0.
    \end{equation}\\
    Compute the advantage by the advantage function, i.e.,
    \begin{equation}
        \hat{A}_i=G_i - V_i, i=0, 1, \ldots, \delta-1.
    \end{equation}\\
    \For{$j=0,1,\ldots, J-1$}{
        $l\leftarrow l+1$. \\
        Compute values similar to Algorithm \ref{training pipeline}, i.e.,
        \begin{align}
             &\left\{\text{Pr}\left(\mathcal{A}_i|\mathcal{S}_i;\theta^a_l\right)\right\} \in \mathbf{Pr}\left(\mathbb{A}|\mathbb{S};\theta^a_l\right) \\
             &\left\{\hat{V}_i\right\} \in \hat{\mathbf{V}}=\pi_{\theta^c_l}\left(\mathbb{S}, \mathbb{A}\right) 
        \end{align}\\
        Update the actor by maximizing the objective \eqref{l_theta}:
        \begin{equation}
            \theta^a_{l+1} = \underset{\theta^a_l,\theta^a_{l_o}}{\arg\!{\max}} \frac{1}{\delta}\sum^{\delta-1}_{i=0}L\left(\theta^a_l,\theta^a_{l_o}\right).
        \end{equation}\\
        Update the critic by minimizing MSE:
        \begin{equation}
            \theta^c_{l+1} = \underset{\theta^c_l,\theta^c_{l_o}}{\arg\!\min} \frac{1}{\delta}\sum^{\delta-1}_{i=0}\left(\hat{V}_i-G_i\right)^2.
        \end{equation}\\
    }
    Clear the buffer.
\end{algorithm}

Moreover, the training algorithm incorporates several techniques to enhance the learning efficiency:
\begin{itemize}
    \item \textit{Exploration Noise}: By introducing randomness, we facilitate policy exploration to uncover and leverage potential advantageous strategies within the environment.
    \item \textit{LSTM for Actor and Critic Networks}: To capture and process long-term dependencies in time-series data, long short term memory (LSTM) networks are employed as the backbone for both actor and critic networks \cite{LSTM}.
    \item \textit{Gradient Clipping}: To prevent gradient explosion during the training process, gradient clipping techniques are applied, ensuring the stability and reliability of the training.
\end{itemize}

\section{Experimental Results and Analysis} \label{section experiment}
This section presents the training and testing datasets, details of parameter settings, and analysis of experimental results. By simulation analysis, we demonstrate the superiority of our proposed RSMA-enabled APVST compared with other video transmission schemes.

\begin{figure*}[tbp]
    \centering
        \subfloat[]{\includegraphics[width=56mm]{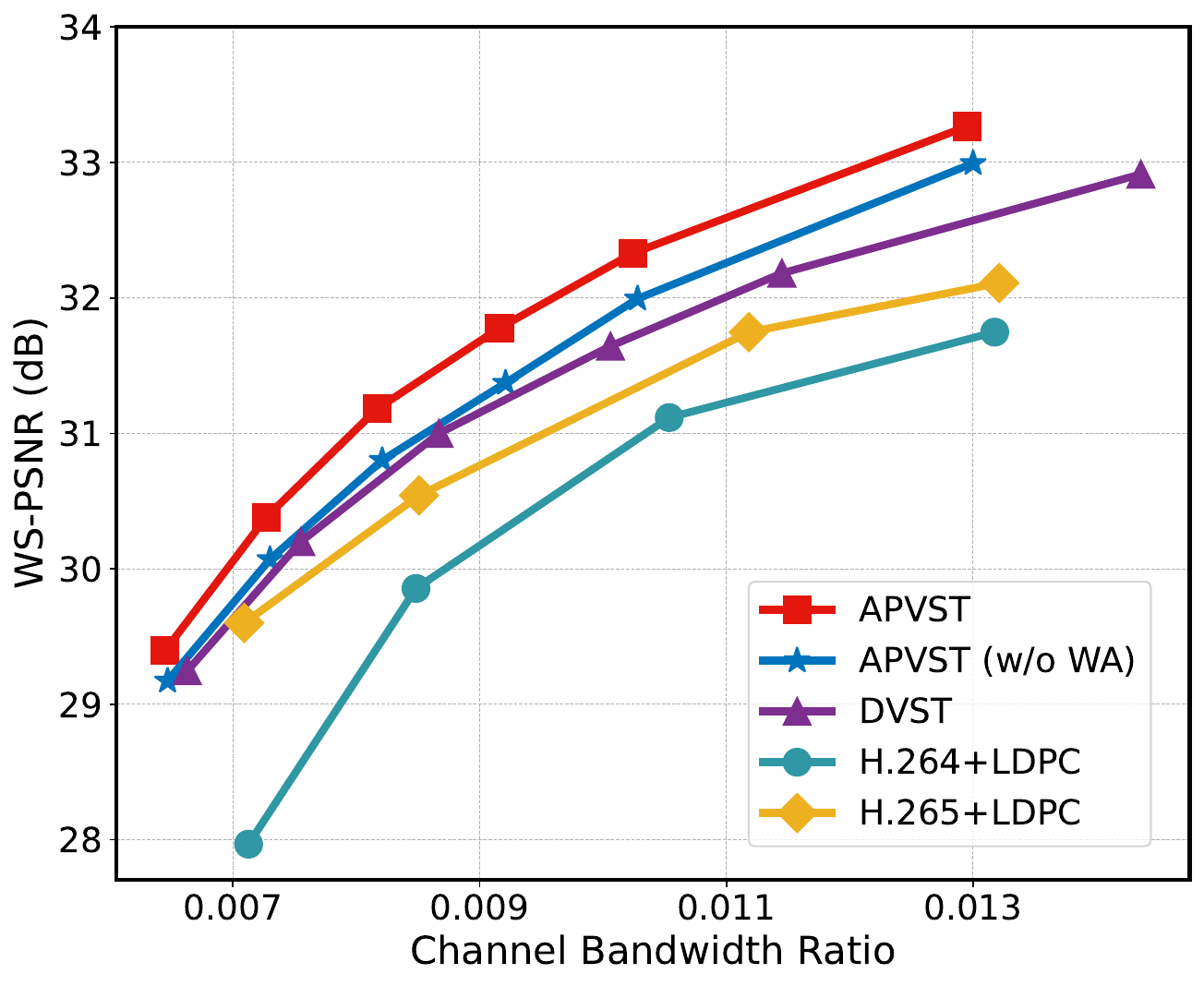} \label{semantic wspnsr vs cbr}}%
    \hfill
        \subfloat[]{\includegraphics[width=56mm]{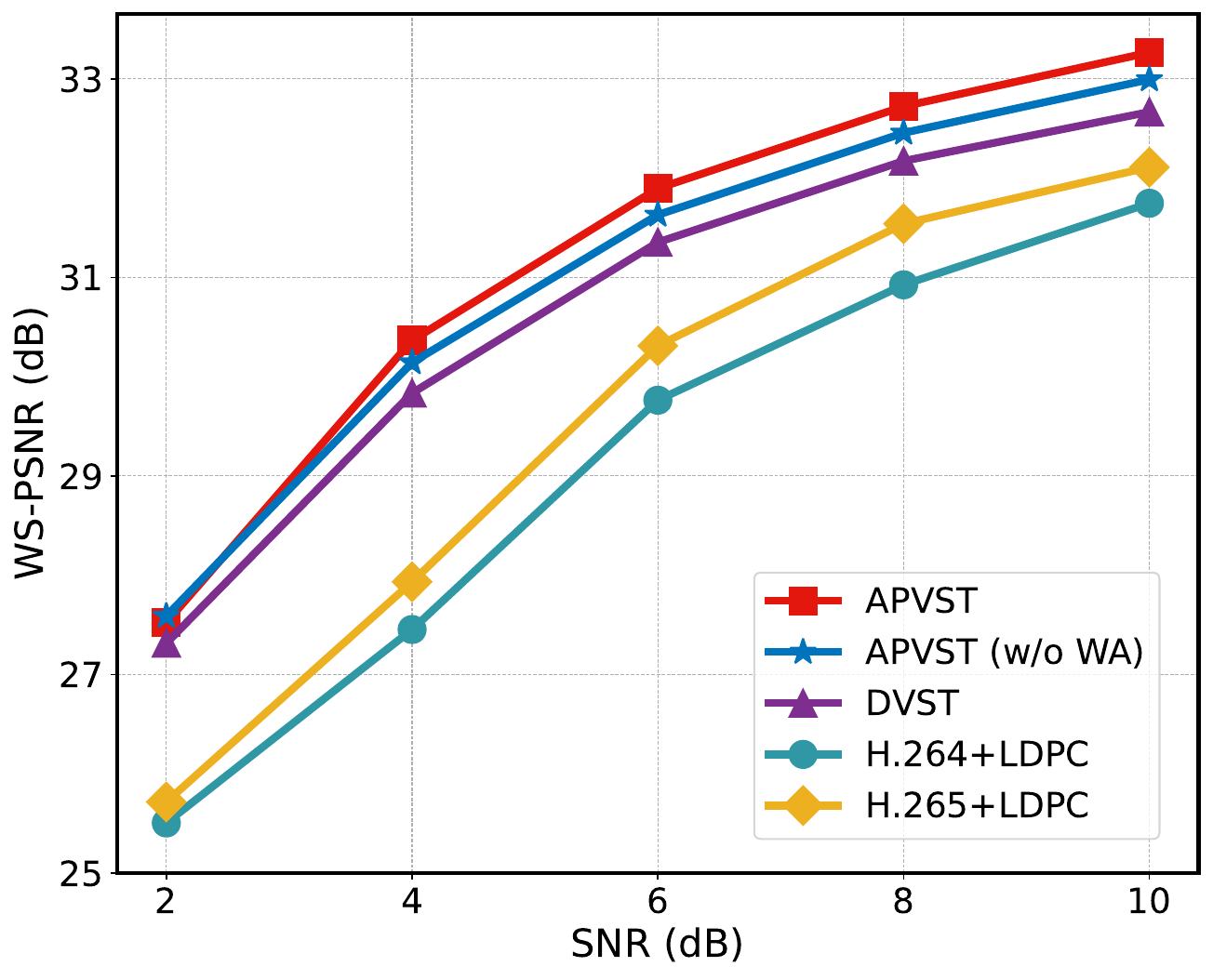} \label{semantic wspnsr vs snr}}
    \hfill
        \subfloat[]{\includegraphics[width=62mm]{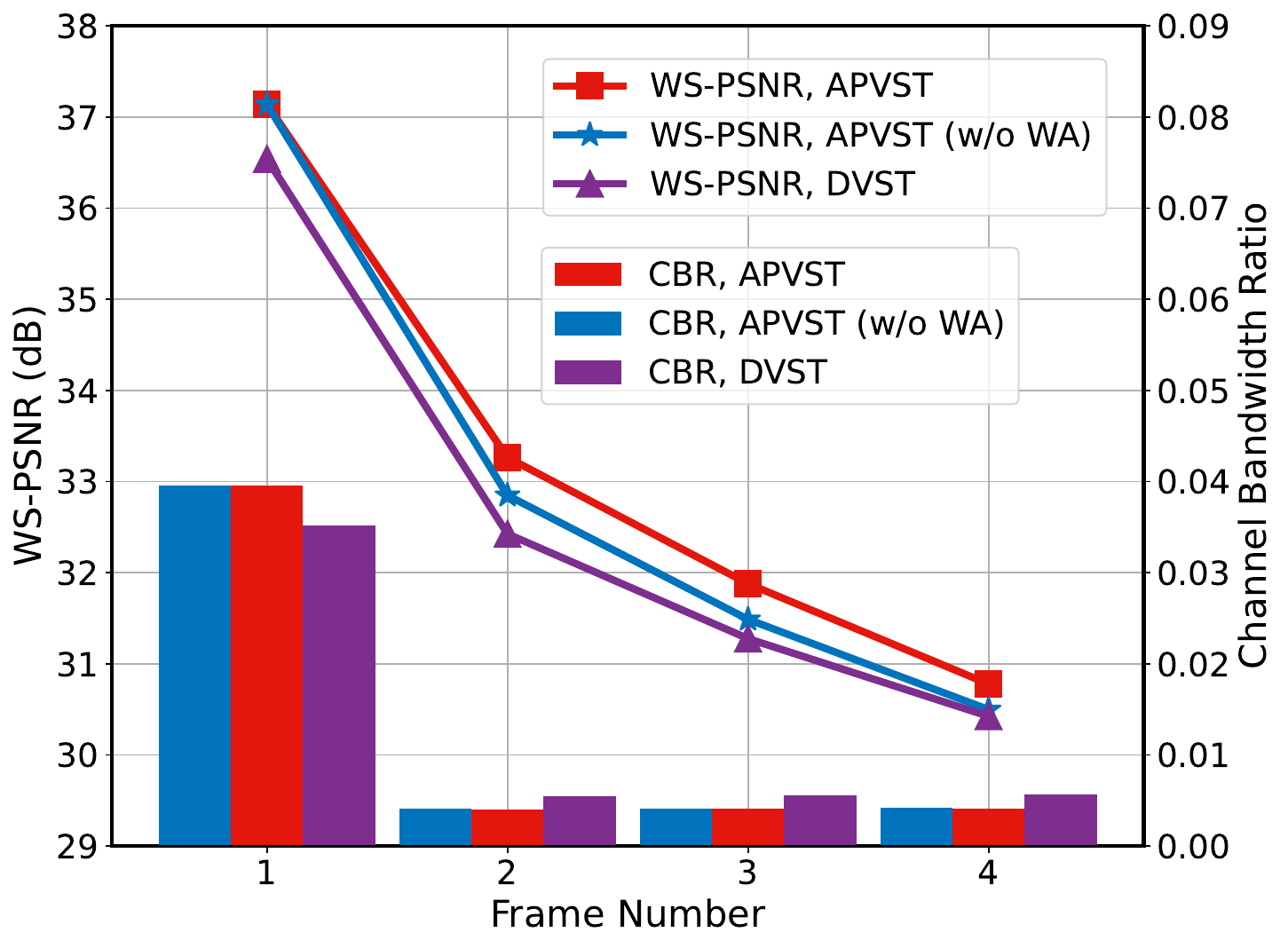} \label{semantic wspnsr vs frame}}
    \caption{(a) WS-PSNR performance vs. CBR, (b) WS-PSNR performance vs. SNR, and (c) WS-PSNR and CBR in a GoP over the AWGN channel at SNR=10dB.}
    \label{semantic wspnsr}
\end{figure*}

\begin{figure*}[tbp]
    \centering
        \subfloat[]{\includegraphics[width=56mm]{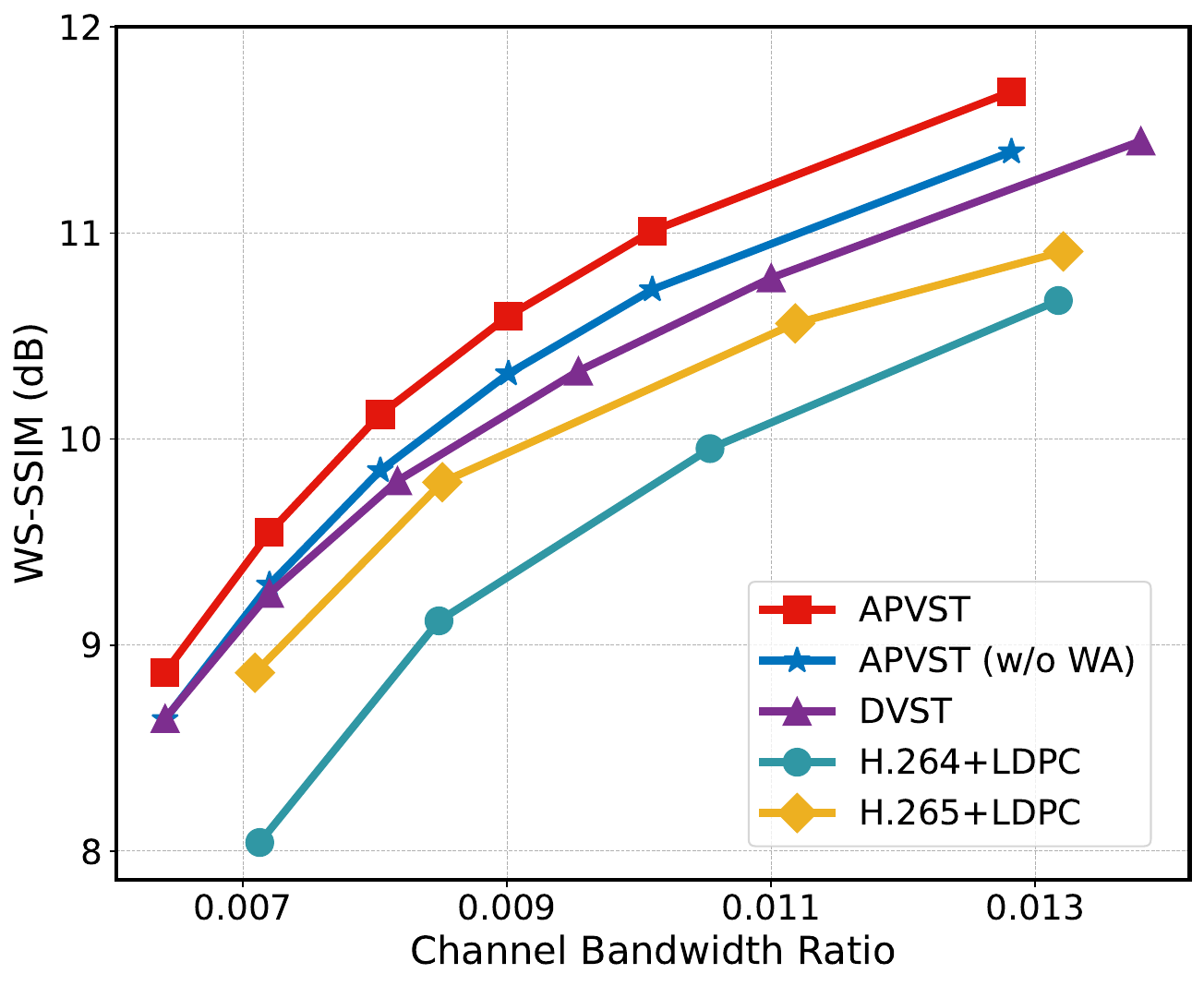} \label{semantic wsssim vs cbr}}%
    \hfill
        \subfloat[]{\includegraphics[width=56mm]{simulation_results/WSPSNRvsSNR.pdf} \label{semantic wsssim vs snr}}%
    \hfill
        \subfloat[]{\includegraphics[width=62mm]{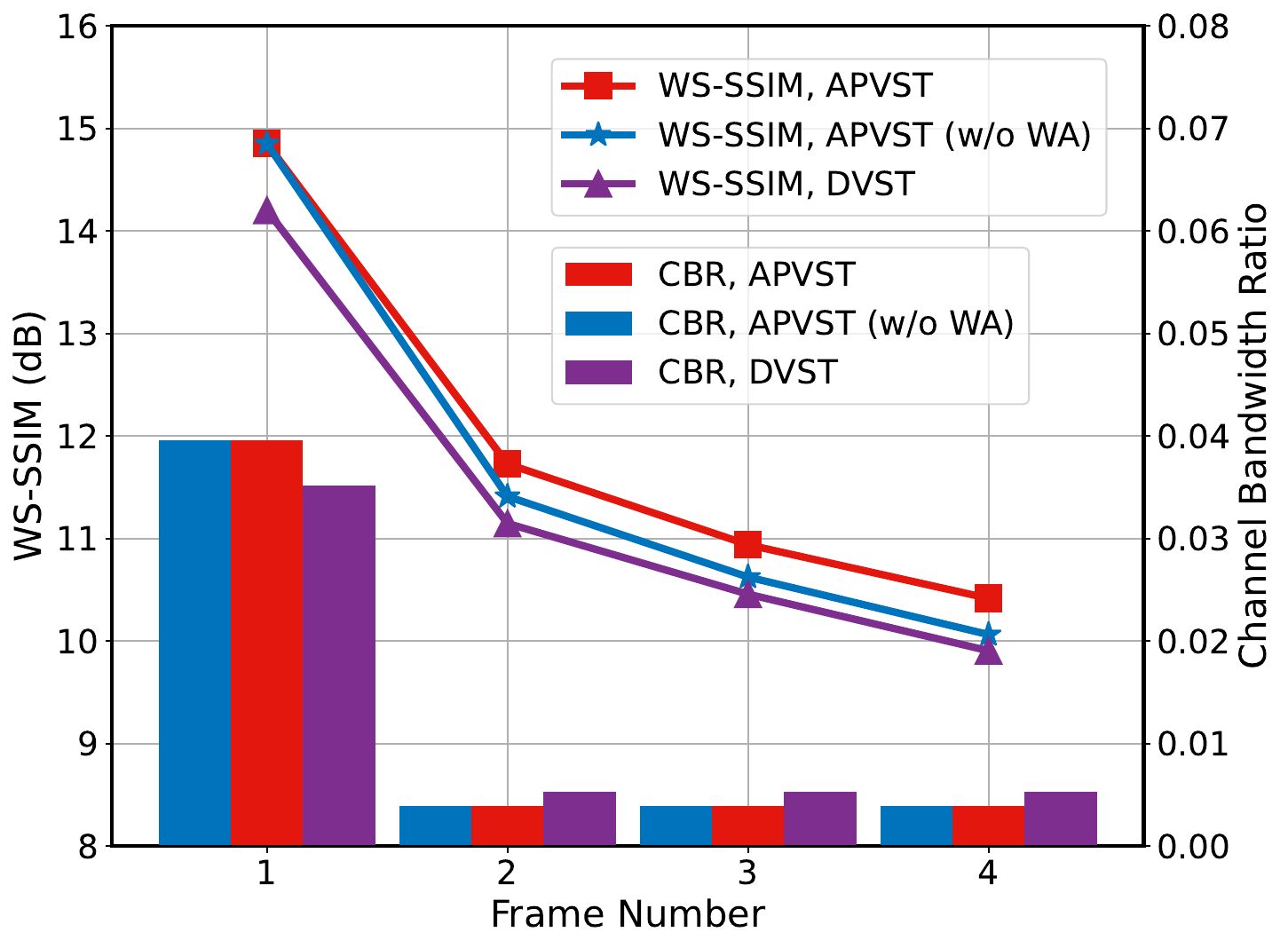} \label{semantic wsssim vs frame}}%
     \caption{(a) WS-SSIM performance vs. CBR, (b) WS-SSIM performance vs. SNR, and (c) WS-SSIM and CBR in a GoP over the AWGN channel at SNR=10dB.}
     \label{semantic wsssim}
\end{figure*}

\begin{table}[h]
    \begin{center}
    \caption{PPO Parameters Settings}
    \label{drl parameters}
    \begin{tabular}{l|l}
        \hline\hline
        \textbf{Parameter} & \textbf{Value} \\
        \hline
        Training epoch & $150$ \\
        Min-batch size & $128$ \\
        Learning rate & $10^{-4}$ \\
        Discount factor $\gamma$ & $0.4$ \\  
        Clipping coefficient $\epsilon$ & $0.1$ \\
        Update iterations $J$ & $2$ \\
        The time span $T$ & $200$ \\
        The timesteps for parameter update $\delta$ & $1000$ \\
        The panoramic frame dims $H\times W$ & $960\times 1920$ \\
        Maximum latency requirement $T_{\max}$ & $10\text{ms}$ \\ 
        Minimum quality requirement $Q_\text{min}$ (WS-PSNR) & $20\text{dB}$ \\
        Maximum quality requirement $Q_{\max}$ (WS-PSNR) & $35\text{dB}$ \\
        Minimum quality requirement $Q_\text{min}$ (WS-SSIM) & $5\text{dB}$ \\
        Maximum quality requirement $Q_{\max}$ (WS-SSIM) & $13\text{dB}$ \\
        \hline\hline
    \end{tabular}
    \end{center}
\end{table}

\subsection{Experimental Environment}
The proposed APVST network is trained on the panoramic video dataset of 360VDS \cite{omnidirectional_video_super_resolution}, the dataset of which contains 590 panoramic videos in ERP. The panoramic frames are resized to $512\times256$ pixels and randomly flipped during training. We evaluate the APVST on the VR scene dataset \cite{gaze_prediction_immersive_videos}, which contains 208 panoramic videos. For the first frame (I-frame) coding of a GoP, we apply nonlinear transform source-channel coding (NTSCC) \cite{ntscc}.

In the APVST experiment, the number of blocks in swin transformerv2 is set to $N_e=N_d=4$. 8 heads and $8\times8$ window size are used in MHSA. The quantization sets in Deep JSCC of the primary and motion links are set as $\mathcal{Q}\!=\!\left\{0,2,4,6,8,10,16,20,26,32,20,48,56,64,80,96\right\}$ and $\mathcal{Q}^\text{ml}\!=\!\left\{0,2,4,6,8,16,32,48\right\}$, respectively. The balance coefficients are set as $\alpha=1/16$ and $\beta=1/16$. In the model training and testing processes, we set the GoP sizes as $N=7$ and $N=4$, respectively. Adam optimizer is adopted as the optimizer. Other parameters are set as: learning rate is ${10}^{-4}$, training batch size is 8, and testing batch size is 1. The whole APVST model was trained on a single A40 GPU for 4 days.

To validate the effectiveness of our proposed APVST scheme in terms of panoramic video transmission, we introduce four schemes for comparison. For semantic communication video transmission schemes, we consider ``APVST (w/o WA)'' and ``DVST'' \cite{DVST}, where ``APVST (w/o WA)'' represents the APVST network without WA module. For traditional video transmission schemes, we consider ``H.264 + LDPC'' and ``H.265 + LDPC'', which employ standard video codecs H.264 and H.265 for source coding and low-density parity check (LDPC) for channel coding.

Based on the training and deployment of the APVST model, further optimization of the actor and critic networks is planned to support the transmission of panoramic video semantic information streams. We employ the VR-HM48 as our training dataset \cite{ppo_train_data} and AVTrack360 as our testing dataset \cite{ppo_test_data}, respectively. Additionally, the FoV for all users is set to a dimension of $60^{\circ}\times 120^{\circ}$. In the RSMA-enabled transmission system, we consider a user set of $\left|\mathcal{U}\right|=6$, scattered within a radius of $R=20m$. The transmission band is set within a frequency band ranging around a center frequency of $f_c=2.6\text{GHz}$ with a bandwidth of $B=200\text{MHz}$. The transmit power spectral density and noise power spectral density are set to $-53 \text{dBm/Hz}$ and $-143 \text{dBm/Hz}$, respectively. Moreover, the parameters during the PPO training are as shown in Table \ref{drl parameters}. To demonstrate the capability of RSMA in maximizing the service quality for users in the transmission of panoramic video semantic streams, we compare it with other multiple access schemes serving as baselines, such as NOMA and OFDMA-based transmission methods.

\begin{figure}[t]
    \centering
    \includegraphics[width=85mm]{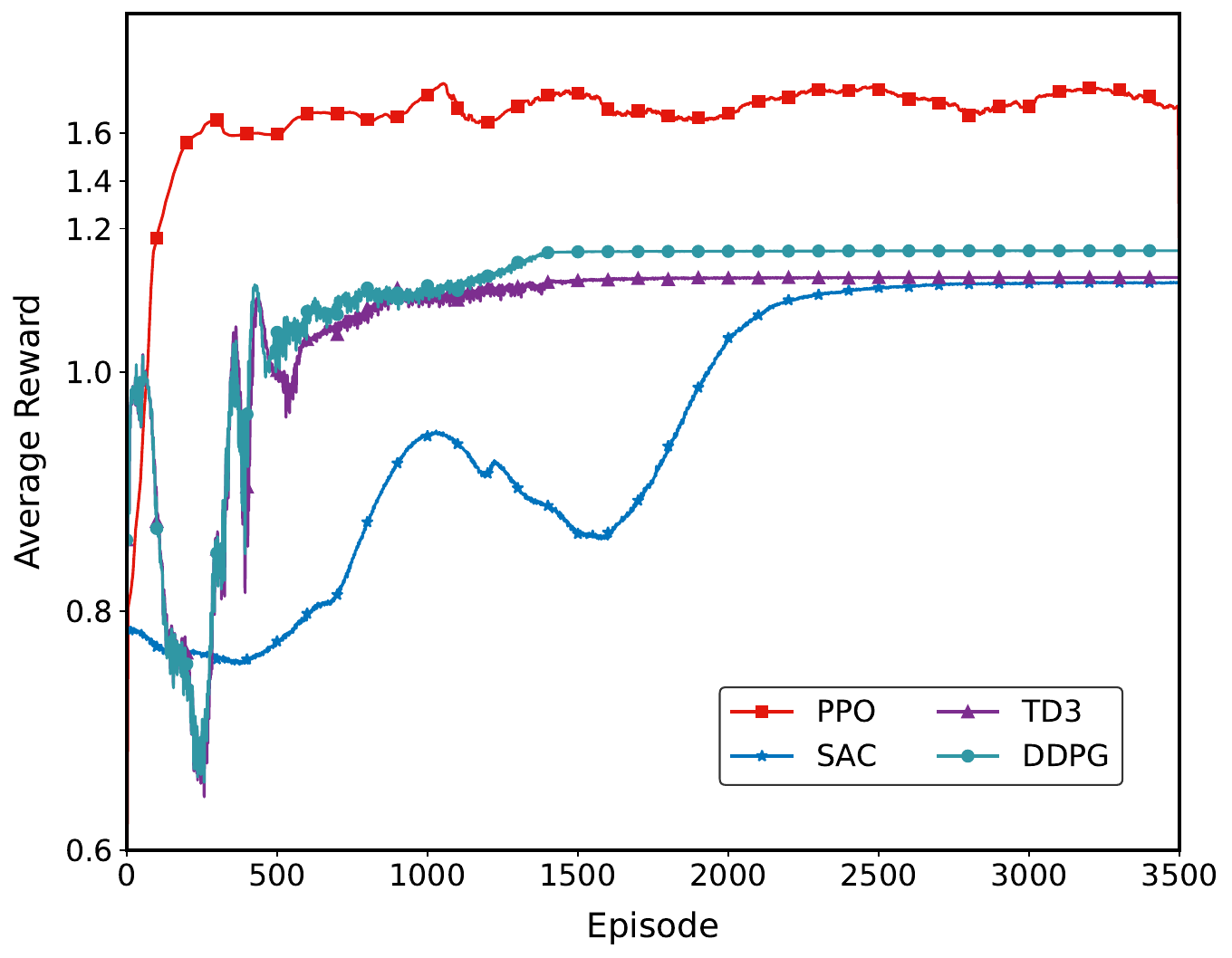}
    \caption{Average rewards vs. episode of PPO, TD3, SAC, and DDPG algorithms.}
    \label{vs different algorithm}
\end{figure}

\subsection{Experimental Analysis}

Fig.~\ref{semantic wspnsr vs cbr} shows WS-PSNR achieved by these five schemes varies with CBR. The test SNR is set as 10dB. Owing to the incorporation of the WA module and the latitude adaptive module, our proposed APVST can save approximately 20\% and 10\% on channel bandwidth cost compared with APVST (w/o WA) and DVST, respectively. Compared with H.264 + LDPC and H.265 + LDPC schemes, the proposed APVST can save bandwidth by about 50\% and 40\%, respectively. These results illustrate the efficiency of our proposed APVST for panoramic video transmission. However, we can see that the WS-PSNR achieved by APVST (w/o WA) and DVST are quite similar under low-CBR conditions. This indicates that the entropy calculated under the guidance of the entropy model and the latitude adaptive network is smaller, and consequently, the impact of $\mathcal{L}^\prime{_t^\mathrm{la}}$ in Equation \eqref{eq19} is also reduced.

Fig.~\ref{semantic wspnsr vs snr} shows WS-PSNR achieved by these five schemes varies with SNR. To ensure fairness, the same CBR is maintained for all schemes. At low SNR levels, semantic transmission significantly outperforms traditional video transmission schemes, which means that the proposed APVST has superior noise immunity properties. However, APVST and DVST achieve the same WS-PSNR at low SNR. This is attributed to the WA module, which primarily learns from semantic feature map $\mathbf{y}_t$ and weight map $\mathbf{w}$ during the training phase, without incorporating noise-related information, thereby reducing the noise immunity of the network. 

Fig.~\ref{semantic wspnsr vs frame} shows WS-PSNR and CBR achieved by the three semantic communication schemes vary with frame number within a GoP. Since the I-frame transmission is without a reference frame, the channel bandwidth ratio is higher compared to the next few frames. The continuous and lossy nature of panoramic video frame transmission leads to the WS-PSNR gradual decrease within a GoP. Compared with DVST and APVST (w/o WA) schemes, the proposed APVST achieves overall gains of about 2.4 dB and 1.1 dB in WS-PSNR under the condition of equal total CBR.

Fig.~\ref{semantic wsssim} shows WS-SSIM achieved by the five schemes varies with CBR and SNR. Similar to the trends observed with WS-PSNR, the proposed APVST scheme can save about 20\% and 40\% on channel bandwidth cost compared with DVST and H.264 + LDPC schemes. This further demonstrates that the APVST can provide a superior immersive experience in panoramic video transmission with less resource consumption.

\begin{figure}[t]
    \centering
    \includegraphics[width=85mm]{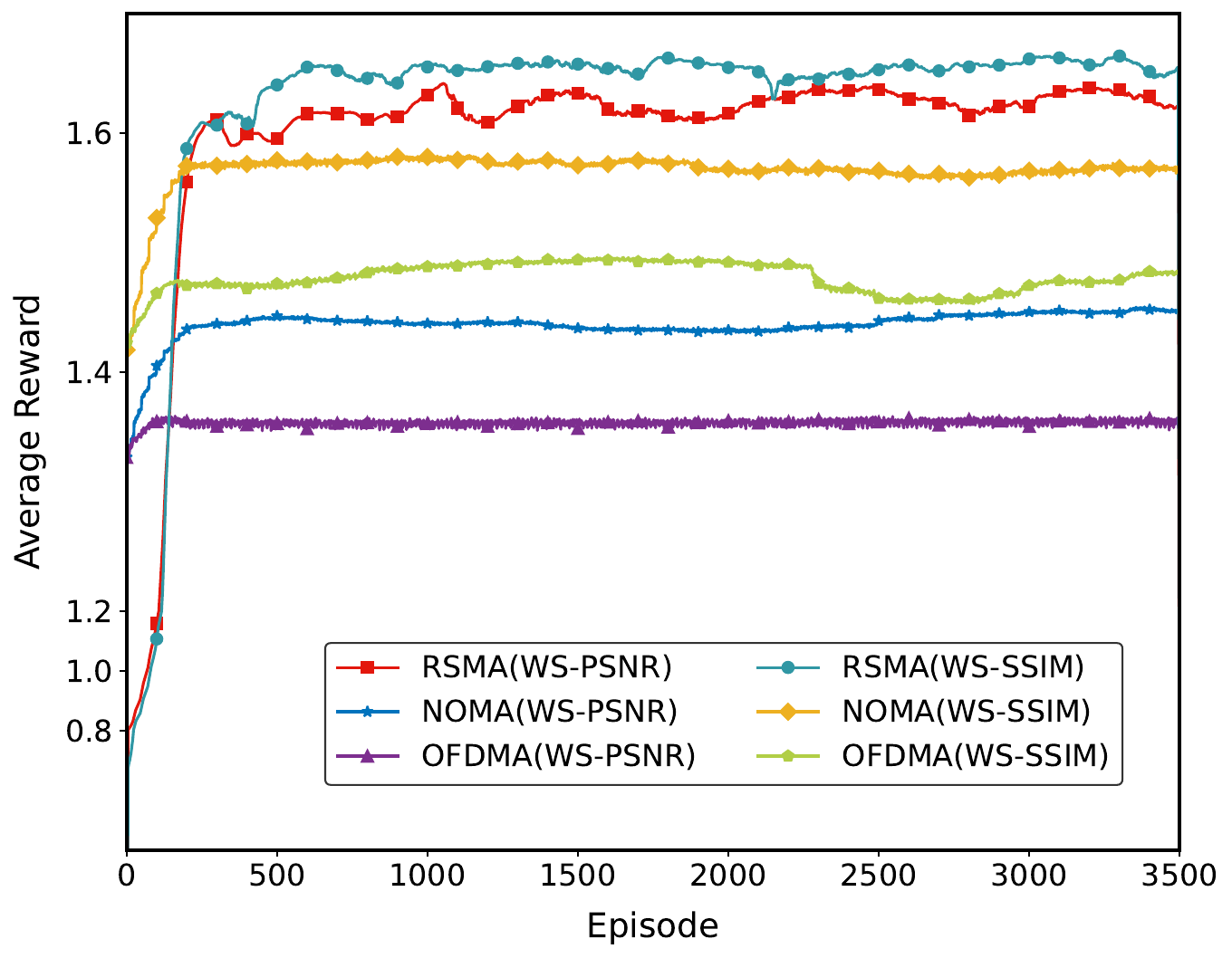}
    \caption{Average rewards vs. episode of RSMA, NOMA, and OFDMA schemes for WS-PSNR and WS-SSIM metrics.}
    \label{PPO-based algorithm 收敛}
\end{figure}

\begin{figure*}[!t]
    \centering
        \subfloat[]{\includegraphics[width=58mm]{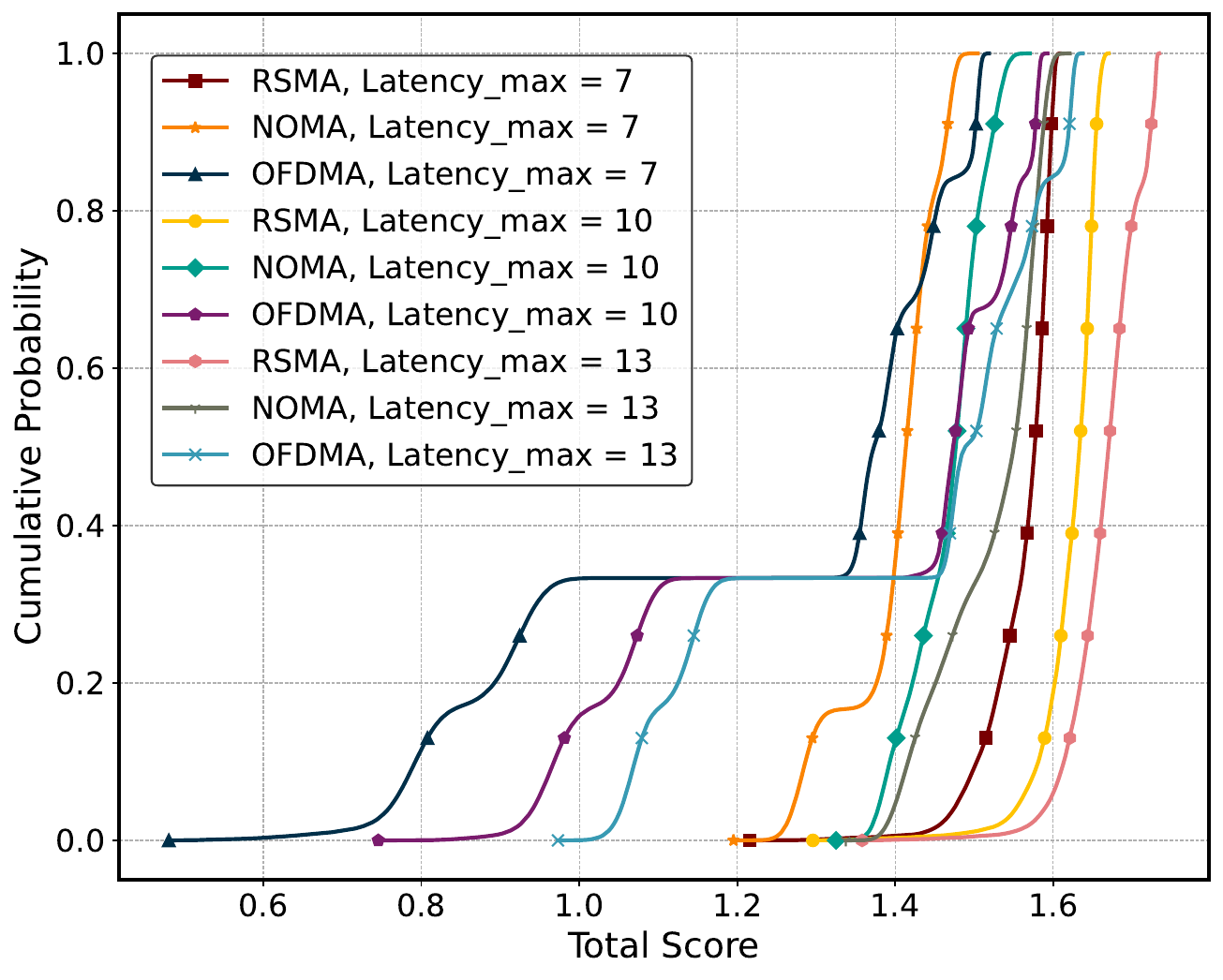} \label{cdf for latency}} 
    \hfill
        \subfloat[]{\includegraphics[width=58mm]{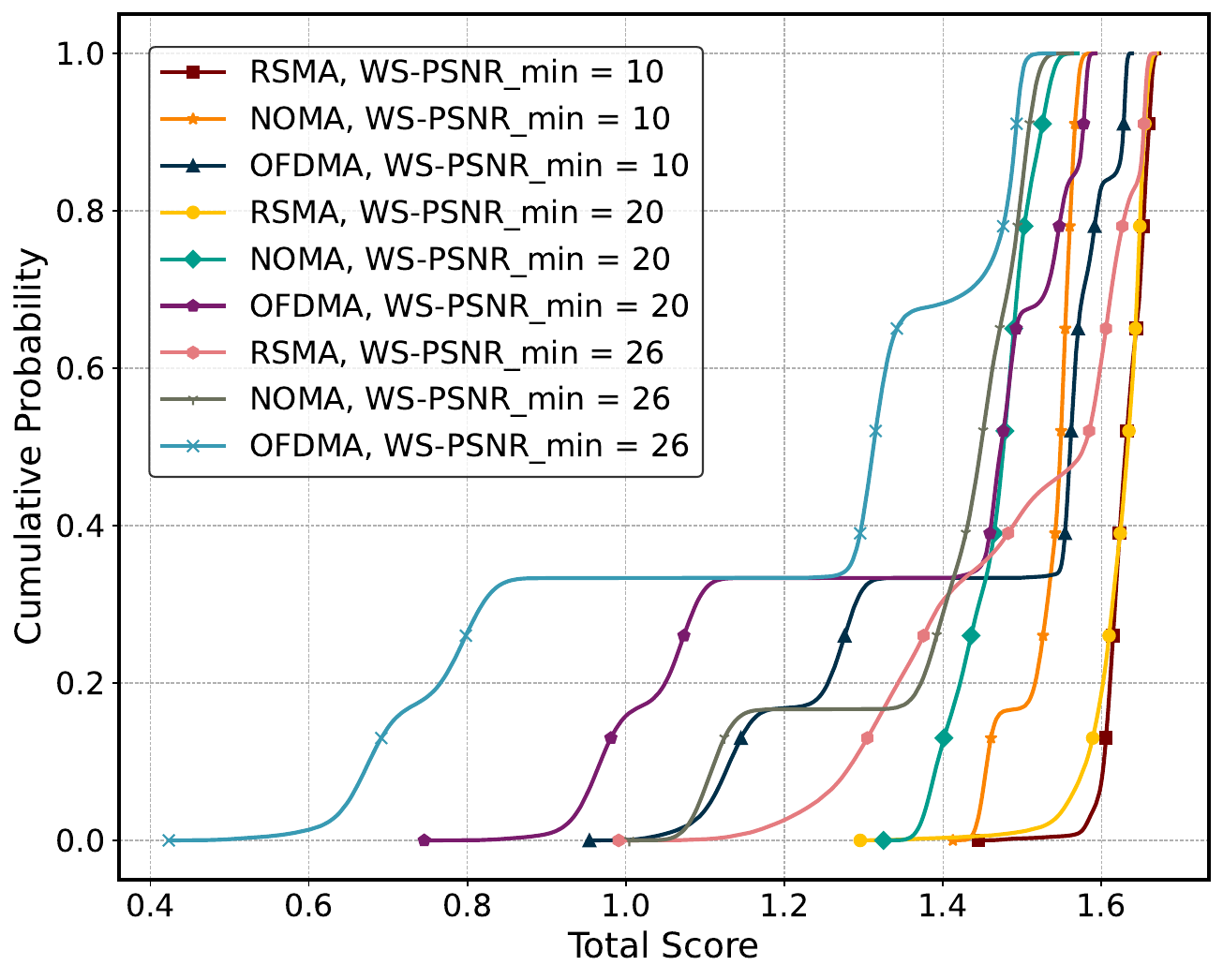} \label{cdf for wspsnr}} 
    \hfill
        \subfloat[]{\includegraphics[width=58mm]{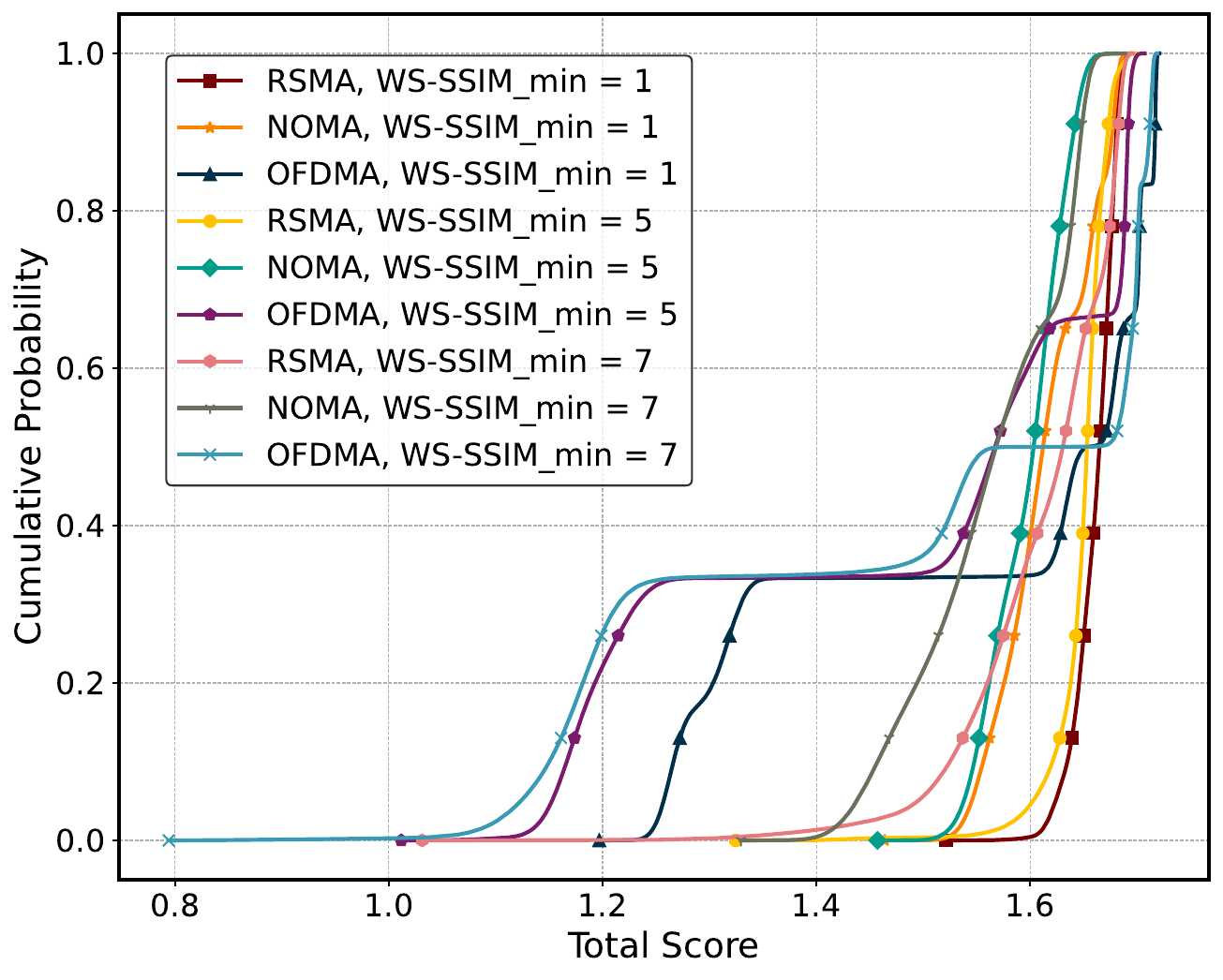} \label{cdf for wsssim}} 
    \caption{CDF curves for users' total score of RSMA, NOMA, and OFDMA under different (a) maximum latency requirements, (b) minimum WS-PSNR requirements, and (c) minimum WS-SSIM requirements.}
    \vspace{-4mm}
    \label{cdf}
\end{figure*}

Next, we will analyze the experimental results of PPO for RSMA-enabled transmission semantic streams. To confirm the superiority of the PPO-based algorithm in efficiently transmitting semantic information in panoramic video streams, we conducted a comparative analysis with three other reinforcement learning algorithms: twin delayed DDPG (TD3) \cite{TD3}, soft actor-critic (SAC) \cite{SAC}, and deep deterministic policy gradient (DDPG) \cite{DDPG}, as illustrated in Fig.~\ref{vs different algorithm}. The results demonstrate that the PPO-based algorithm quickly reaches a higher reward level and maintains stability early in the experiments, showcasing outstanding convergence speed and stability. In contrast, although the SAC algorithm experiences significant fluctuations initially, it gradually exhibits sustained performance improvement, eventually stabilizing. The TD3 and DDPG algorithms, however, show slower performance growth and more considerable fluctuations throughout the training process. Therefore, due to its excellent rapid convergence capabilities and high stability, the PPO-based algorithm is confirmed as the optimal choice for addressing this issue.

We also compared the performance of three technologies RSMA, NOMA, and OFDMA in transmitting semantic streams for panoramic videos. The results, depicted in Fig.~\ref{PPO-based algorithm 收敛}, indicate that our proposed RSMA-enabled APVST scheme significantly outperforms the other two techniques. RSMA not only demonstrates faster convergence speeds and higher average rewards across both WS-PSNR and WS-SSIM metrics, but also maintains superior performance throughout the training process. Specifically, for the WS-PSNR metric, RSMA-enabled APVST scheme achieves performance gains of 13\% and 20\% over NOMA and OFDMA, respectively. For the WS-SSIM metric, the gains are 6\% and 12\%. These results underscore the efficiency and stability of RSMA in handling the complex demands of multi-user panoramic video transmission.

To further validate the efficiency of RSMA technology in the transmission of panoramic videos, Fig.~\ref{cdf} shows the cumulative distribution function (CDF) curves of total scores for all users. Overall, the RSMA curves are positioned to the right of those for NOMA and OFDMA, indicating that a greater proportion of users achieve higher total scores under RSMA support. Specifically, Fig.~\ref{cdf for latency} demonstrates that as the maximum latency limit is increased, the number of users meeting this criterion rises, correspondingly shifting the CDF curve to the right. Conversely, reducing the latency limit moves the CDF curve to the left. Similarly, Fig.~\ref{cdf for wspsnr} and ~\ref{cdf for wsssim} illustrate the impact on total scores when adjusting the minimum threshold for immersive experience quality. Increasing the minimum requirement for immersive quality results in fewer users meeting the criterion, shifting the CDF curve to the left. Conversely, lowering this standard shifts the curve to the right. These visuals effectively demonstrate the potential of RSMA technology to enhance user experience quality and reduce latency.

\begin{figure*}[!t]
    \centering
        \subfloat[]{\includegraphics[width=57mm]{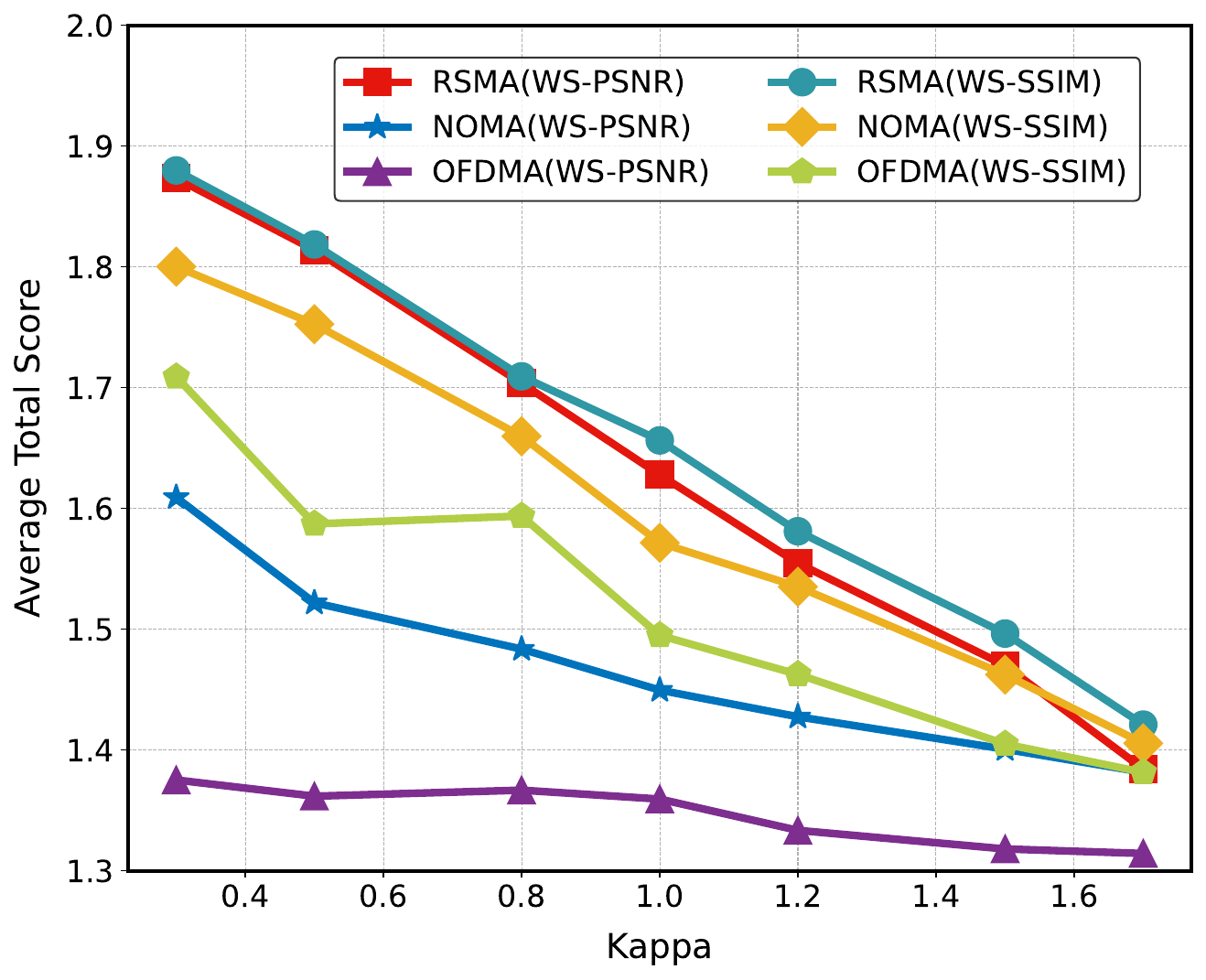} \label{different weight for rewrad}} 
    \hfill
        \subfloat[]{\includegraphics[width=58mm]{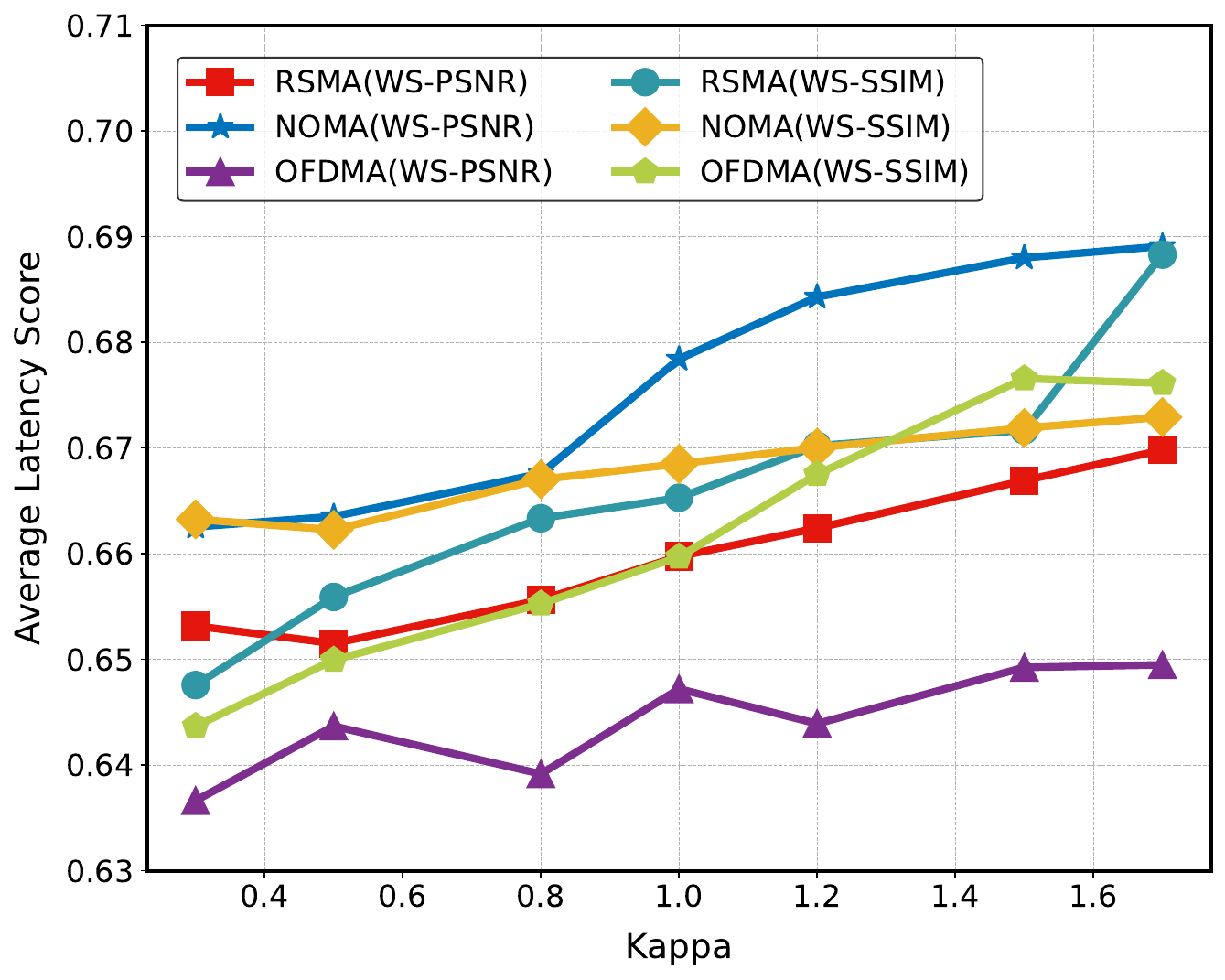} \label{different weight for latency}} 
    \hfill
        \subfloat[]{\includegraphics[width=57mm]{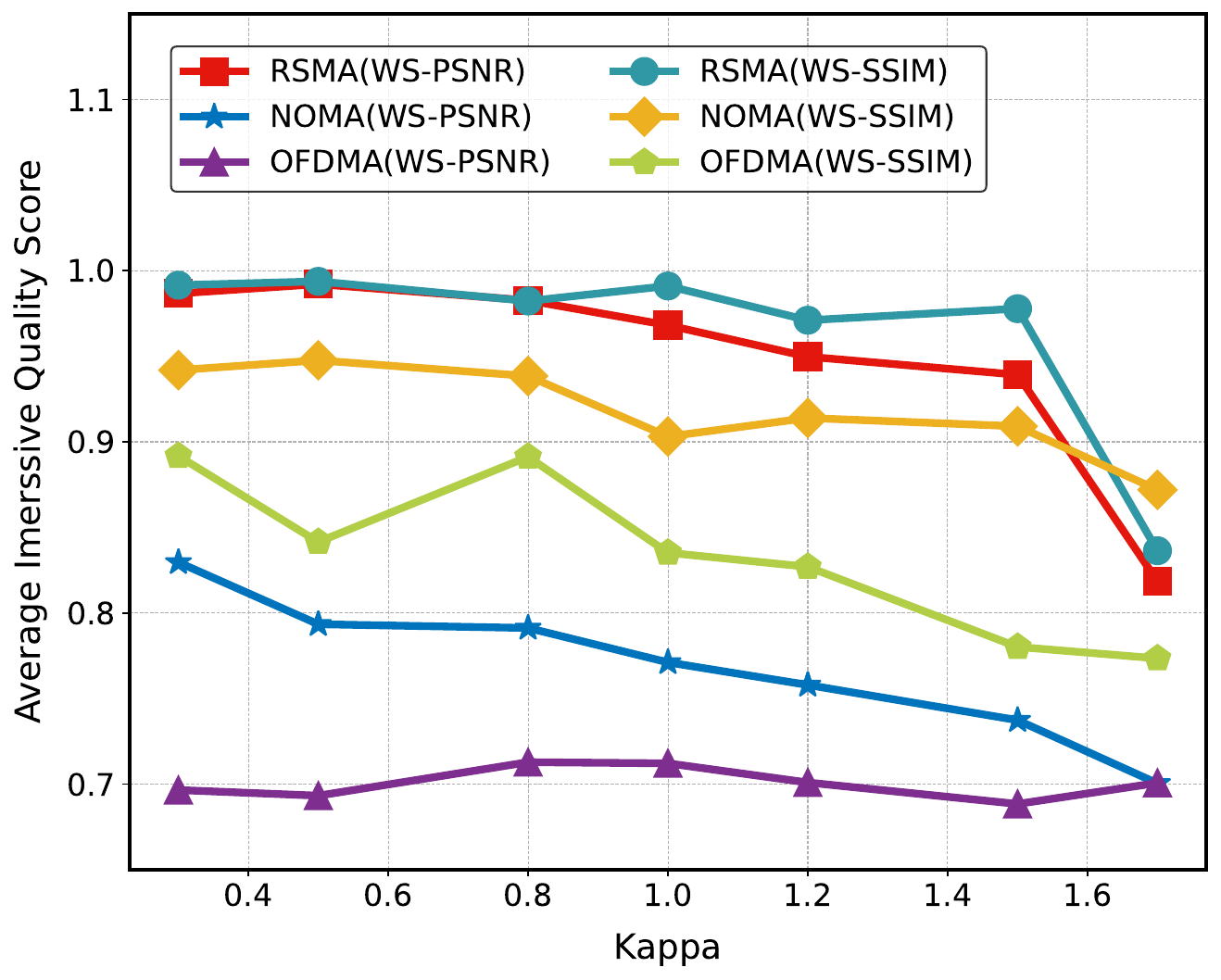} \label{different weight for immersive}} 
    \caption{(a) Average total score, (b) Average latency score, and (c) Average immersive quality score vs. $\kappa$ of RSMA, NOMA, and OFDMA for WS-PSNR and WS-SSIM metrics.}
    \vspace{-4mm}
    \label{different weight}
\end{figure*}

Fig.~\ref{different weight} illustrates the impact of varying the balance coefficient $\kappa$ on the optimization outcomes. Specifically, increasing the weight of the transmission delay fraction (i.e., increasing the $\kappa$ value) caters to scenarios with stringent requirements on transmission delays, while reducing the $\kappa$ value correspondingly decreases the optimized transmission delay score. For scenarios demanding higher immersive experience quality, lowering the $\kappa$ value enhances the weight of the immersive experience quality score, thus user experience can be optimized. However, as observed from \eqref{latency to score}, the mapping from transmission delay to score is nonlinear. When \( T_{\max} \) is low, it becomes challenging for the transmission delay score to reach higher levels, leading to an imbalance between the two scores. For instance, the maximum transmission delay score is 0.69, while the maximum immersive experience quality score approaches 1. As shown in Fig.~\ref{different weight for latency}, increasing $\kappa$ results in a downward trend in the average total score, primarily due to the relatively low transmission delay scores.

\begin{figure*}[!t]
    \centering
        \subfloat[]{\includegraphics[width=58mm]{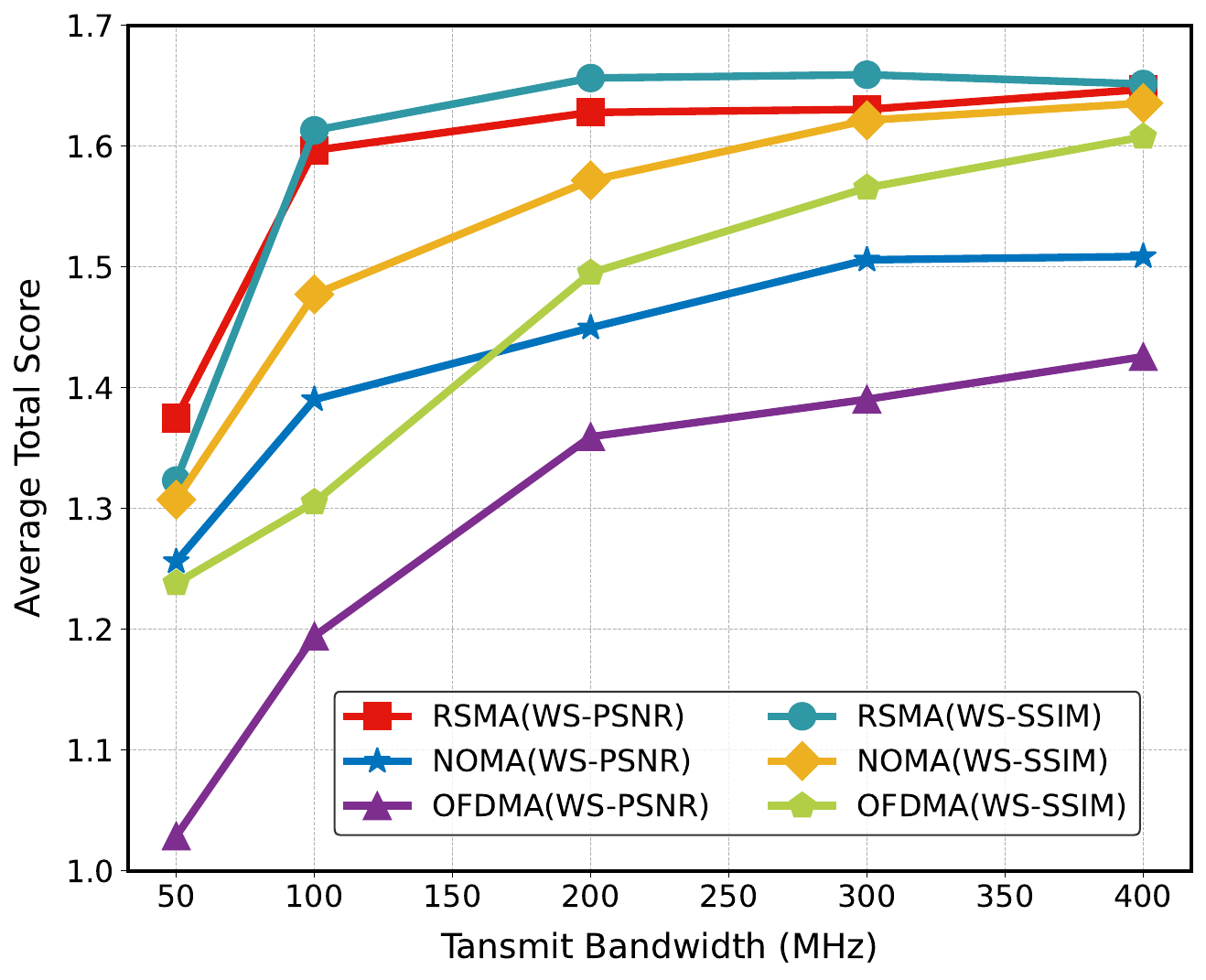} \label{different bandwidth reward}}%
    \hfill
        \subfloat[]{\includegraphics[width=58mm]{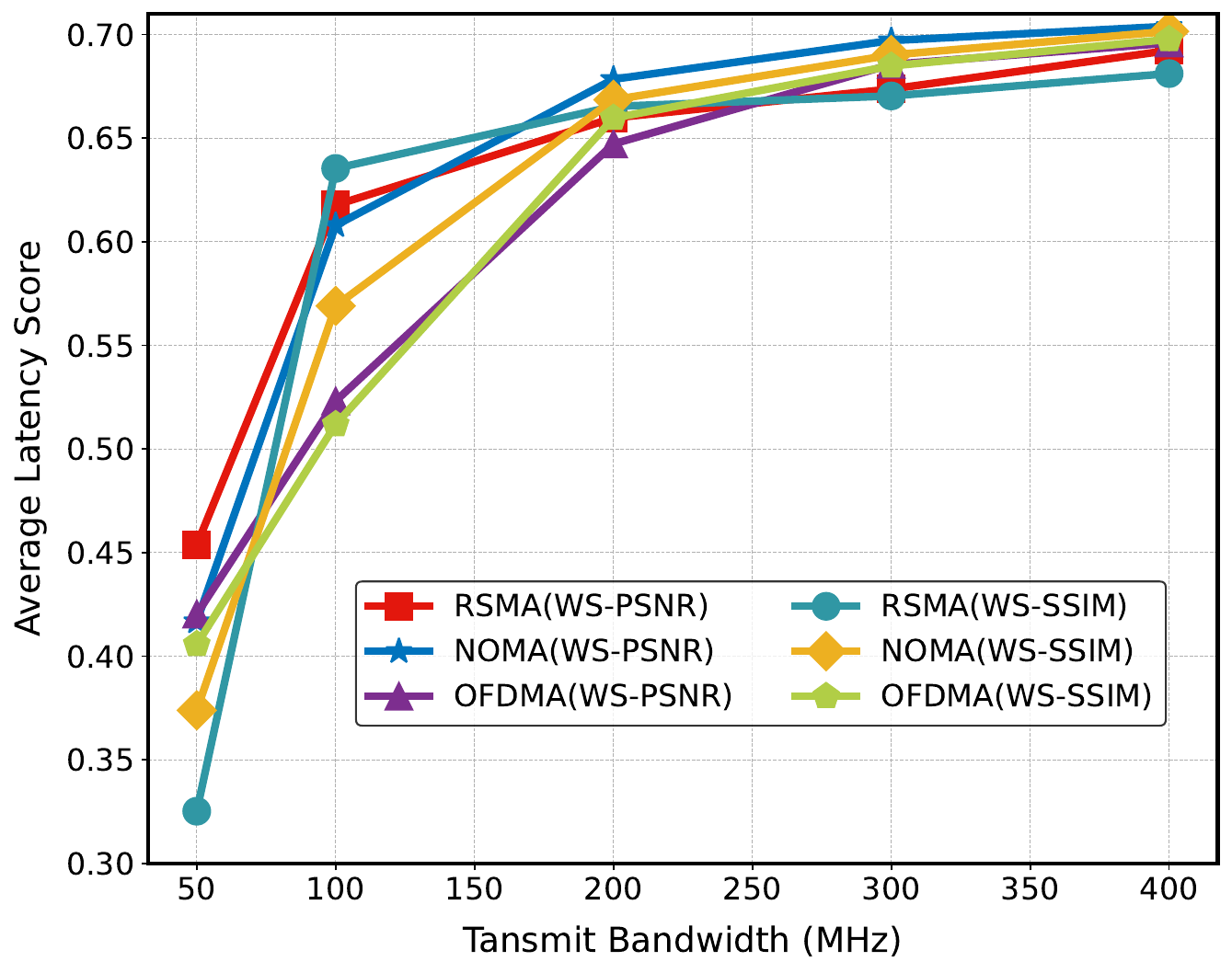} \label{different bandwidth latency}}
    \hfill
        \subfloat[]{\includegraphics[width=58mm]{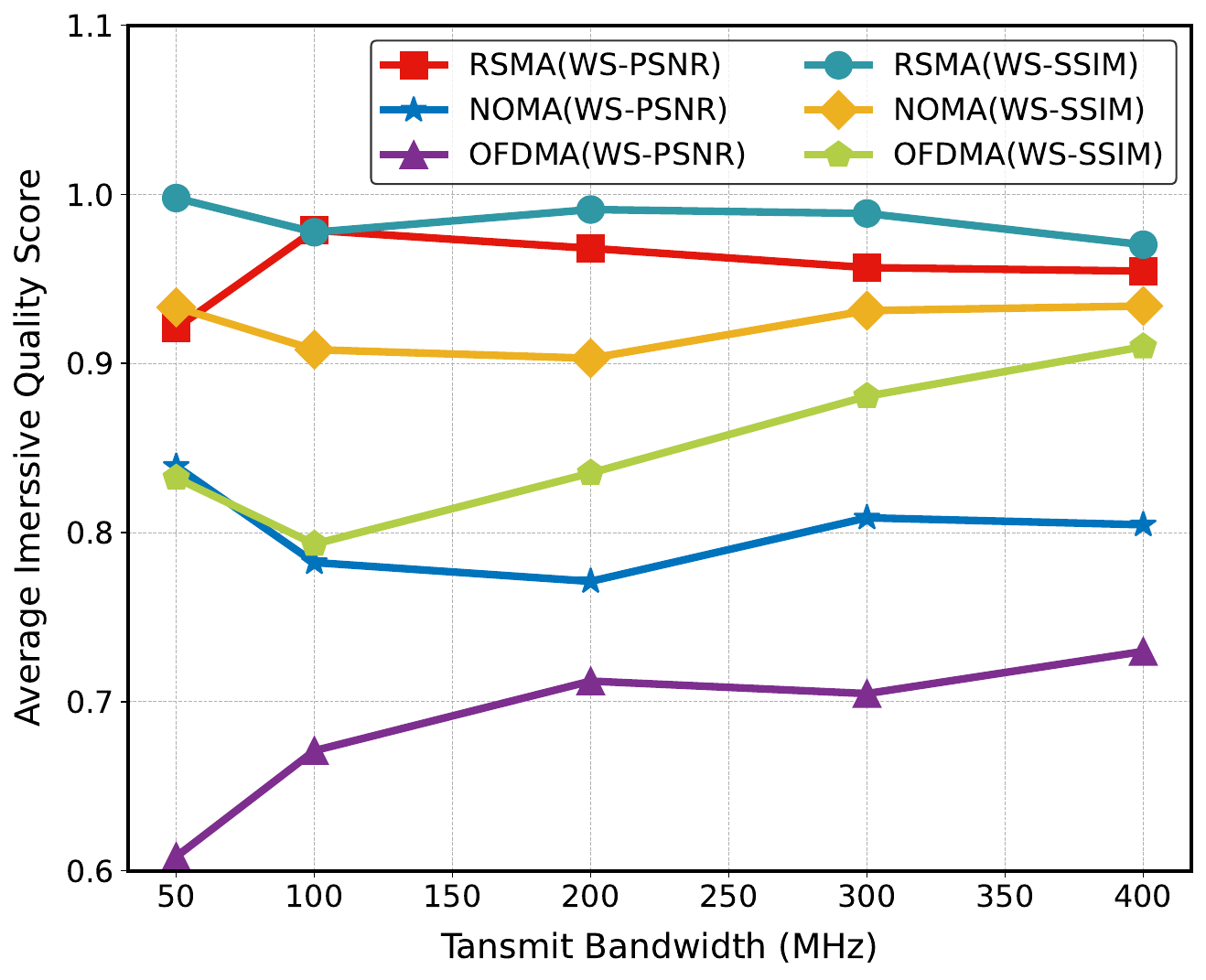} \label{different bandwidth immersive}}%
    \caption{(a) Average total score, (b) Average latency score, and (c) Average immersive quality score vs. bandwidth of RSMA, NOMA, and OFDMA for WS-PSNR and WS-SSIM metrics.}
    \vspace{-2mm}
    \label{different bandwidth}
\end{figure*}

In Fig.~\ref{different bandwidth}, we adjust transmission bandwidth $B$ to evaluate the performance of different technologies in transmitting panoramic video content, which also impacts the maximum transmission power \(P_{\max}\) of BS. Fig.~\ref{different bandwidth reward} demonstrates that the RSMA technology significantly outperforms other technologies in terms of average total scores. Specifically, under the same bandwidth conditions, according to the WSS-PSNR metric, RSMA's average total score is approximately 11\% and 24\% higher than NOMA and OFDMA, respectively. For the WSS-SSIM metric, the achieved score by RSMA-enabled APVST scheme is even more impressive, surpassing NOMA and OFDMA by 4\% and 10\%, respectively. This significant performance enhancement can be attributed to the increased channel capacity and reduced transmission delays that come with increased bandwidth. As shown in Fig.~\ref{different bandwidth latency} and ~\ref{different bandwidth immersive}, the transmission delay scores rapidly increase with increasing bandwidth. However, once the bandwidth reaches a certain level, the rate of improvement in delay scores gradually slows. This indicates that while bandwidth expansion can effectively reduce delays within a certain range, beyond a certain threshold, additional increases in bandwidth contribute less to further reductions in delay. Additionally, changes in bandwidth have a negligible impact on immersive experience quality scores.
\section{Conclusion} \label{conclusion}

This paper proposes an adaptive panoramic video semantic transmission framework based on RSMA. The proposed framework, APVST, utilizes a Deep JSCC structure and attention mechanism to adaptively extract and encode semantic features from panoramic frames, effectively enhancing spectral efficiency and saving bandwidth. By integrating an entropy model and a dimension-adaptive module, APVST achieves more precise rate control. Furthermore, the proposed weighted self-attention module enhances the quality of the immersive experience for users. Additionally, the framework takes into account the overlap in the FoV when users watch panoramic videos, utilizing RSMA to effectively split the required panoramic video semantic streams into common and private messages. We formulate this transmission process as a joint optimization problem of latency and immersive experience quality, and exploit the PPO-based algorithm to find the optimal solution, aiming to maximize users' QoS scores. Extensive simulation results demonstrate that our APVST framework significantly outperforms other semantic and traditional video transmission schemes. Moreover, compared to NOMA and OFDMA, RSMA-enabled APVST scheme achieves higher performance gains in the transmission of panoramic videos, confirming the efficiency and superiority of RSMA in handling high-demand video content. These results not only showcase the application potential of RSMA and semantic communication in immersive communication environments but also provide a solid foundation and valuable references for further research.

\footnotesize
\bibliographystyle{IEEEtran}

\end{document}